\documentclass[twocolumn]{aastex631}

\newcommand{\red}{\textcolor{black}}
\newcommand{\blue}{\textcolor{black}}

\newcommand{\xmm}{\textit{XMM-Newton} }
\usepackage{cancel}
\usepackage{tikz}
\usepackage{booktabs}
\usepackage{enumitem}
\usepackage{multirow}
\usepackage{array}
\usepackage{subfigure}
\usepackage{soul}
\usepackage{makecell}
\usepackage{amsmath}
\nocite{*}

\defcitealias{henley2012xmm}{HS12}
\defcitealias{Qu:2023aa}{Paper II}

\begin{document}
\title{The \xmm Line Emission Analysis Program (X-LEAP) I: Emission Line Survey of \ion{O}{7}, \ion{O}{8}, and Fe L-Shell Transitions}

\correspondingauthor{Zhijie Qu}
\email{quzhijie@uchicago.edu}

\author[0000-0002-9943-1155]{Zeyang Pan}
\affiliation{Key Laboratory of Optical Astronomy, National Astronomical Observatories, Chinese Academy of Sciences, 20A Datun Road, Beijing, 100101, People's Republic of China}
\affiliation{School of Astronomy and Space Science, University of Chinese Academy of Sciences, Beijing, 100049, People's Republic of China}

\author[0000-0002-2941-646X]{Zhijie Qu}
\affiliation{
Department of Astronomy $\&$ Astrophysics, The University of Chicago, Chicago, IL 60637, USA}

\author[0000-0001-6276-9526]{Joel N. Bregman}
\affiliation{Department of Astronomy, University of Michigan, Ann Arbor, MI 48109, USA}

\author{Jifeng Liu}
\affiliation{Key Laboratory of Optical Astronomy, National Astronomical Observatories, Chinese Academy of Sciences, 20A Datun Road, Beijing, 100101, People's Republic of China}
\affiliation{School of Astronomy and Space Science, University of Chinese Academy of Sciences, Beijing, 100049, People's Republic of China}
\affiliation{Institute for Frontiers in Astronomy and Astrophysics, Beijing Normal University, Beijing, 102206, People's Republic of China}
\affiliation{New Cornerstone Science Laboratory, National Astronomical Observatories, Chinese Academy of Sciences, Beijing, 100012, People's Republic of China}

\begin{abstract}
The \xmm Line Emission Analysis Program (X-LEAP) is designed to study diffuse X-ray emissions from the Milky Way (MW) hot gas, as well as emissions from the foreground solar wind charge exchange (SWCX). 
This paper reports an all-sky survey of spectral feature intensities corresponding to the \ion{O}{7}, \ion{O}{8}, and iron L-shell (Fe-L) emissions. 
These intensities are derived from 5418 selected \xmm observations with long exposure times and minimal contamination from point or extended sources. 
For 90\% of the measured intensities, the values are within $\approx$ 2--18 photons cm$^{-2}$ s$^{-1}$ sr$^{-1}$ (line unit; L.U.), $\approx$ 0--8 L.U., and $\approx$ 0--9 L.U., respectively.
We report long-term variations in \ion{O}{7} and \ion{O}{8} intensities over 22 years, closely correlating with the solar cycle and attributed to SWCX emissions.
These variations contribute $\sim30\%$ and $\sim20\%$ to the observed intensities on average and peak at $\approx$ 4 L.U. and $\approx$ 1 L.U. during solar maxima.
We also find evidence of short-term and spatial variations in SWCX, indicating the need for a more refined SWCX model in future studies.
In addition, we present SWCX- and absorption-corrected all-sky maps for a better view of the MW hot gas emission. These maps show a gradual decrease in oxygen intensity moving away from the Galactic center and a concentration of Fe-L intensity in the Galactic bubbles and disk.
\end{abstract}

\section{Introduction}
\label{sec:intro}

The circumgalactic medium (CGM) is a diffuse, multi-phase gas surrounding the galaxy. It plays an important role in the formation, evolution, and interactions of galaxies \citep[see the reviews][]{Tumlinson:2017aa, Donahue:2022aa}.
In particular, the CGM encodes the feedback materials, metals, and energy ejected from the galaxy disk, as well as the accreted gas from the intergalactic medium (IGM).
Studying the properties and distribution of the CGM provides valuable constraints on these uncertain processes.

Over the last two decades, the multi-phase CGM in the Milky Way (MW) has been extensively explored using multi-wavelength observations \citep[see the review][]{Putman:2012aa}.
X-ray observations reveal that the hot CGM ($T\gtrsim 10^6$ K) is a massive component in the multi-phase medium \citep[e.g.,][]{Snowden:1995aa, henley2010xmm, henley2012xmm}. 
\blue{Notably, the stars and interstellar medium (ISM) in the MW only account for $\approx$ 30--50\% of the baryons, expected from the cosmic average baryonic fraction \citep[e.g.,][]{sommer2006missing, McGaugh:2010aa}. This discrepancy suggests a deficit of $\approx 3 \times 10^{11} M_{\odot}$ \citep{Anderson:2010aa}, positioning the hot CGM as a principal candidate for the ``missing baryons".}

The hot gaseous phase is primarily probed through the emission or absorption lines of highly ionized metals, such as the \ion{O}{7} K$\alpha$ and \ion{O}{8} K$\alpha$.
These two ions exist at temperatures of $\approx$ 1--$2 \times 10^6$ K assuming collisional ionization equilibrium (CIE), which is consistent with the expected virial temperature of the MW.
Extensive efforts have been made to estimate the total hot gas mass using the oxygen emission lines \citep[e.g.,][]{Gupta:2012aa, Miller:2013aa, miller2015constraining, Nakashima:2018aa, kaaret2020disk}.
However, these estimates yield considerable uncertainty, ranging from $\approx$ 3--$4\times 10^{10} M_{\odot}$ to 1--$2 \times 10^{11} M_{\odot}$ \citep{miller2015constraining, faerman2017massive}.

\blue{This uncertainty primarily arises from two factors. 
First, the emission or absorption column is affected by plasma density, temperature, and metallicity together.
The emission line intensity scales 
as $I \propto n_e^{2} \epsilon (T) Z$, while the absorption column scales as $N\propto n_e f(T) Z$, where $\epsilon $ and $f$ are the emissivity and ionization fraction, respectively.}
Accurate assessments of temperature and metallicity are crucial to reliably infer the hot gas density from X-ray observations.

Second, the MW hot gas emission is contaminated by foreground emissions from the solar wind charge exchange (SWCX) and the Local Hot Bubble (LHB). The SWCX occurs when solar wind ions exchange charges with neutral atoms in the Solar System. Its intensity varies temporally and spatially \citep[see][for a review]{kuntz2019solar}, contributing a large amount to the diffuse soft X-ray background (SXRB) at energies below 1 keV \citep[e.g.,][]{2000ApJ...532L.153C, koutroumpa2009solar, uprety2016solar}. The LHB is thought to be an irregular ``local cavity" extending $\sim$ 100 pc from the Sun \citep{lallement20033d}. Although the origin of the LHB has many explanations \citep[e.g.,][]{maiz2001origin, linsky2021could, Zucker:2022}, its gas temperature is measured to be T $\approx 10^6$ K \citep[e.g.,][]{mccammon1990soft, snowden2000catalog, liu2016structure, yeung2023srg}, a temperature high enough to emit X-rays \citep{1977ApJ...217L..87S}. Shadowing studies toward nearby molecular clouds indicate that the LHB contributes negligibly to the observed \ion{O}{8} emission but accounts for part of the \ion{O}{7} emission \citep[e.g.,][]{smith2007suzaku,koutroumpa:2011}.

The \xmm archive offers unique opportunities to overcome these challenges with observations over two decades.
It provides a deep, sensitive, and spatially resolved view of the X-ray sky, and also allows us to consider long-term temporal variations associated with solar activity. The high sensitivity enables us to detect faint emissions due to the hot gas.
The spatial resolution enables the removal of contamination due to other X-ray sources, improving the accuracy of the line intensity measurements. 

We design the \xmm Line Emission Analysis Program (X-LEAP) aiming to study the diffuse X-ray emissions from the foreground SWCX and the MW hot gas.
In this first paper, we conduct a line emission survey, which measures three specific spectral features in the SXRB: \ion{O}{7} K$\alpha$ ($\approx$ 0.57 keV), \ion{O}{8} K$\alpha$ ($\approx$ 0.65 keV), and transitions due to the iron L-shell (Fe-L; $\approx$ 0.8 keV). They are referred to as emission lines for simplicity hereafter, although they are combinations of individual lines.
These lines are the strongest in both SWCX and MW hot gas emissions. The oxygen lines serve as primary tracers of the hot gas, while the Fe-L line is a frequently detected feature in SXRB spectra peaking in emissivity at $\approx 7 \times 10^6$ K. This temperature is above MW's virial temperature, suggesting that the Fe-L feature may trace an even hotter phase of the MW gas. 
Recent CGM studies show evidence of such hotter component with a super-virial temperature $\approx 10^7$ K, associated with the Fe-L line \citep[e.g.,][] {das2019multiple, bluem2022widespread, ponti2022abundance, bhattacharyya2023hot}. The origin of this component is debated, with no consensus among these studies. Including this feature in our survey is crucial to further explore and understand its properties and origin.

Because of the complexities of measuring these lines, we adopt a line extraction method introduced by \citet[][hereafter \citetalias{henley2012xmm}]{henley2012xmm}.  
This method does not estimate temperature or emission measure but extracts line intensities directly from the spectra.
This method is insensitive to assumptions of the underlying physical model of the hot gas, SWCX, and other background emissions, reducing measurement uncertainties.
This is particularly important because the SXRB includes contributions from multiple emission sources that are difficult to separate. 

Here, we present the line measures of the \ion{O}{7}, \ion{O}{8}, and Fe-L lines from the 22-year \xmm archive data, along with initial findings, including intensity fluctuations in foreground SWCX and all-sky intensity distributions of the MW.

\label{sec:DataReduction}

\begin{figure*}[t]
\centering
\includegraphics[width=0.8\paperwidth]{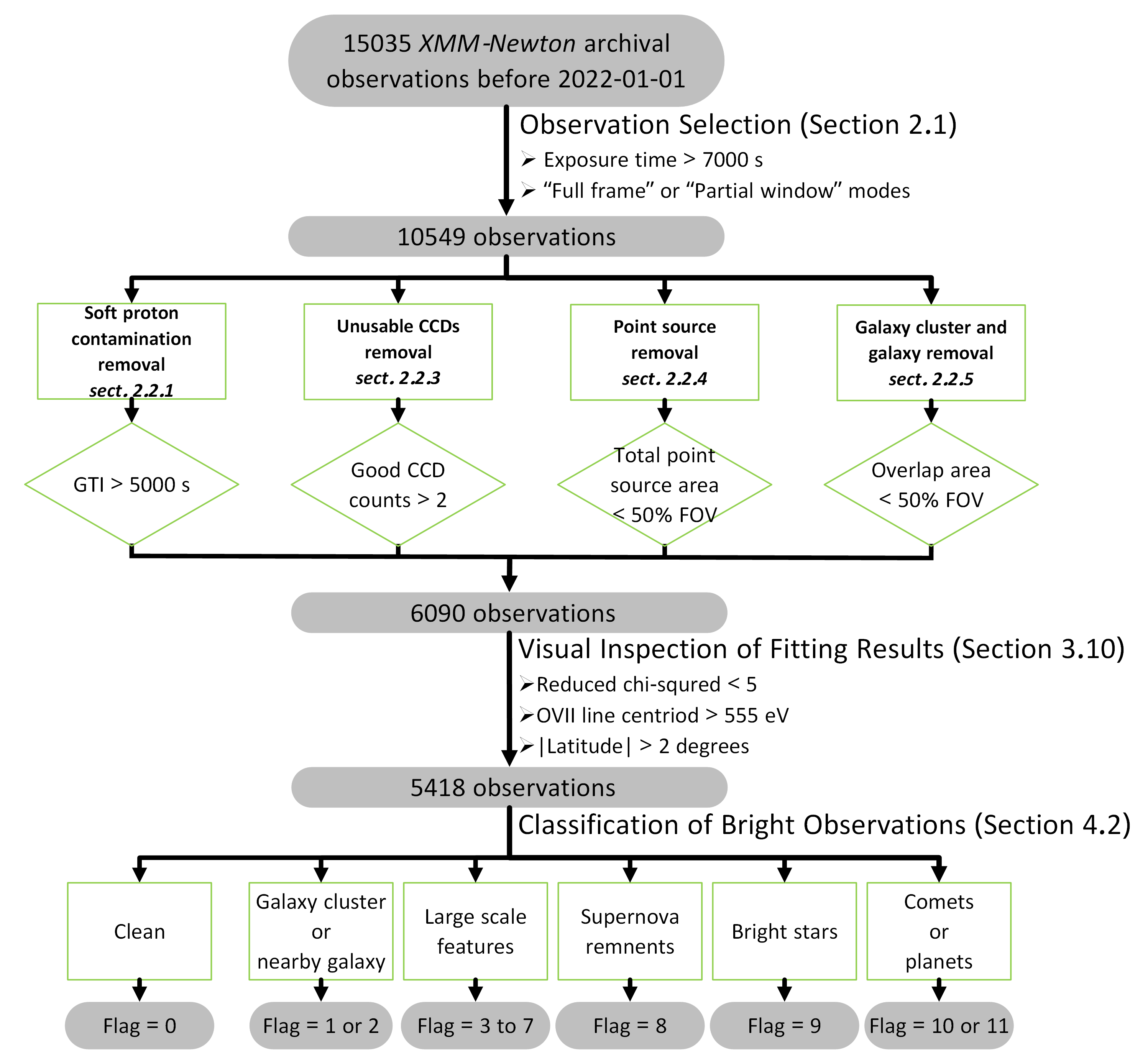}
\caption{Flowchart illustrating the data reduction process and observation classification. Explanations of the flags are in Table \ref{tab:flags}.}

\label{flowchart}
\end{figure*}

The paper is structured as follows: 
Section \ref{sec:DataReduction} outlines the data selection and reduction process. 
The emission components in the SXRB and our approach to modeling and fitting are introduced in Section \ref{sec:COMPONENTS AND SPECTRAL MODEL}.
In Section \ref{sec:Results}, we present a catalog of observed \ion{O}{7}, \ion{O}{8}, and Fe-L intensities. We also discuss the distribution and correlations of these intensities. Additionally, we compare our intensity measurements with those from \citetalias{henley2012xmm}.
Our approach to characterizing long-term SWCX variation is described in Section \ref{sec:SWCX_variation}.
Section \ref{sec:Soft X-ray distribution of the MW} introduces methods for SWCX- and absorption-corrections, and presents the corrected all-sky maps. 
Section \ref{sec:Discussion} discusses further implications of the sample, including the use of \ion{O}{7} line centroids as a diagnostic of solar activity, the evidence of magnetospheric SWCX, and the spatial dependence of SWCX on ecliptic latitude. We also discuss the forthcoming works for this program.
Finally, Section \ref{sec:summary} summarizes the key findings.

\section{Data Selection and Reduction}

We present a new survey of \ion{O}{7}, \ion{O}{8}, and Fe-L emission features in the SXRB using the \xmm archive data before 2022 January 1st. From 15,035 total observations, our data selection and reduction processes yield 6155 data points across 6090 observations, with some observations containing multiple exposures. The dataset is further reduced to 5418 observations, based on the quality check of spectral fitting in Section \ref{subsec: Visual Inspection}. These observations are optimized to study the soft X-ray emissions of the MW hot gas and the SWCX.
In particular, they have relatively long exposure times and low contamination due to soft proton backgrounds, external galaxies, and galaxy clusters.
The selection criteria and data processing are summarized in Figure \ref{flowchart} and detailed in this section.

\subsection{Initial Observation Selection}
In the X-LEAP program, we only consider data obtained from the EPIC-MOS detectors (MOS1 and MOS2) because of their large field of view (FOV, $\approx 30'$) and high sensitivity at energies of the interested lines.
\blue{We consider only observations with EPIC-MOS exposure times longer than 7 ks to ensure a reasonable signal-to-noise ratio, enabling reliable detection of faint emission.}
In addition, we only consider observations in ``Full Frame" or ``Partial Window" modes. These modes observe less bright and/or time-varying X-ray sources compared to other modes and read out all seven or most CCDs (CCD\#2-7). This ensures full or near-full FOV coverage, making them more suitable for diffuse emission studies. Consequently, 4486 observations are ignored from the above criteria.

\subsection{Data Processing}
\label{DP}

The data processing starts with the observation data files (ODFs) obtained from the \xmm Science Archive\footnote{http://nxsa.esac.esa.int/nxsa-web/\#tap}. For each ODF, the calibrated photon event files for both EPIC-MOS detectors are generated using the XMM-SAS \texttt{emchain} script. To improve data quality, public software tools are used, including HEAsoft version 6.30.1\footnote{https://heasarc.gsfc.nasa.gov/docs/software/heasoft/}, SAS version 20.0.0\footnote{https://www.cosmos.esa.int/web/xmm-newton/sas} \citep{gabriel2004xmm}, \xmm Extended Source Analysis Software\footnote{http://heasarc.gsfc.nasa.gov/docs/xmm/xmmhp\_xmmesas.html} (XMM-ESAS; \citealt{kuntz2008epic}; \citealt{snowden2011cookbook}) and \texttt{FTOOLS}\footnote{ http://heasarc.gsfc.nasa.gov/ftools} \citep{blackburn1995asp}.
The following subsections outline the detailed steps adopted to clean and optimize these calibrated photon event files.
   
\subsubsection{Soft Proton Background}
\label{pep}
The soft proton background (SPB) arises from interactions between the EPIC-MOS detectors and soft protons (SPs), observable as flares in the photon count rate.
These flares can last for seconds to hours and significantly contaminate observed data \citep{snowden2004xmm}. 
To minimize this contamination, we utilize the XMM-SAS \texttt{espfilt} script.
This script identifies the flares as time intervals with count rates 2.5$\sigma$ above the median, assuming a Gaussian distribution of the observed photon count rates. 
The resulting Good Time Interval (GTI) file is used to generate the SP-filtered event file.
We only include observations with at least one exposure having a GTI $>5$ ks in the following analysis.
This flare-filtering process cannot remove all SP contamination \citep{henley2010xmm}.
The remaining SPB contamination is modeled as a power law component in the subsequent spectral analysis (Section \ref{subsec:Residual SP cotamination}; see also \citealt{kuntz2008epic}).
The flare-filtering process removes 942 observations from the final sample.

\subsubsection{The Quiescent Particle Background}
\label{QPB}
The quiescent particle background (QPB) is induced when high-energy particles interact with the detectors and the surrounding instruments. It includes both continuum and instrumental line emissions.
The continuum and weak lines are calculated for each observation using the XMM-ESAS \texttt{mos\_back} script \citep{kuntz2008epic}, and then subtracted to obtain the QPB-cleaned SXRB spectrum. 

Besides the modeled QPB spectrum, the QPB contains two strong instrumental lines: Al K$\alpha$ and Si K$\alpha$, located at 1.49 keV and 1.74 keV, respectively.
These lines cannot be well-modeled by the QPB spectrum. They are represented as two Gaussian lines in the spectral model, detailed in Section \ref{subsec:InstrumentalLines}.

\subsubsection{Unusable CCD Removal}
Among all seven CCD chips in each EPIC-MOS detector, some chips may occasionally be exceptionally bright or completely nonfunctional \citep{kuntz2008epic}.
In particular, MOS1 CCD\#3 is damaged and no longer in use due to a micrometeorite strike on 2012 December 11th.
Such anomalous CCD chips are identified using the XMM-SAS \texttt{emanom} script. \blue{It computes the count rate ratio between the 2.5--5.0 keV and 0.4--0.8 keV energy bands for each functional chip, and} compares to a predefined threshold. Chips with ratios above this threshold are classified as ``good".  Conversely, nonfunctional chips, unable to generate photon counts, are classified as ``off" \citep{kuntz2008epic}. 
Our analysis only considers exposures with at least three ``good" CCD chips, ensuring coverage of $>40\%$ of the FOV. 
In total, 21 observations are excluded due to insufficient CCD chips.

\begin{figure*}[t!]
\centering
\label{GCcheese}\includegraphics[width=0.833\paperwidth]{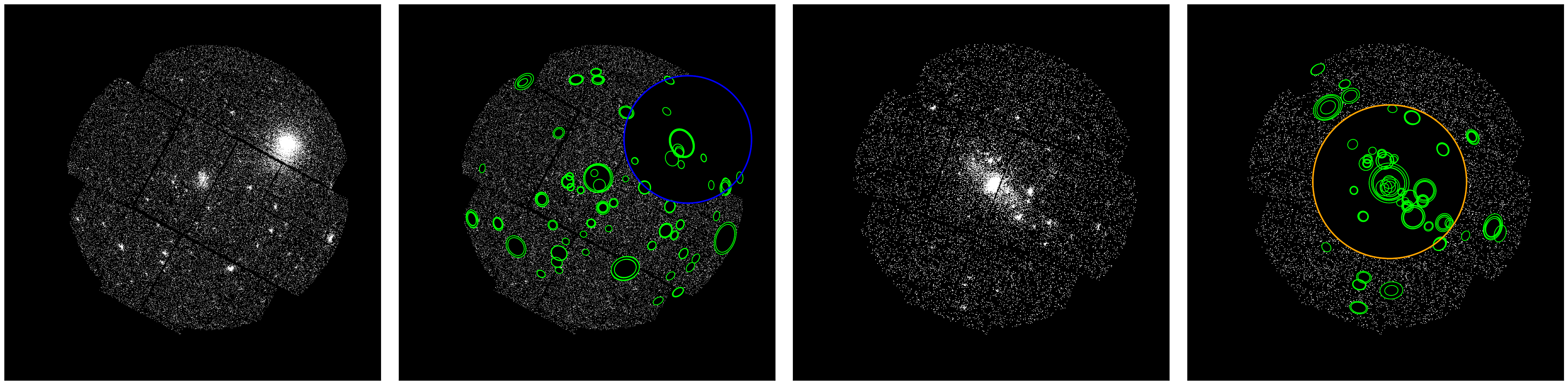}
\caption{Examples of point source and extended source removal on EPIC-MOS2 images. The sequential panels depict images from ObsIDs: 0675010901 (the left two panels) and 0304850901 (the right two panels). Each set presents the image before and after the removal of point sources (green ellipses) and either a galaxy cluster (Abell 3718 in blue) or a nearby galaxy (NGC 253 in orange). See Section \ref{EXS} for size determination of the extended sources.}
\label{fig:cheese}
\end{figure*}

\subsubsection{Point Source Removal}
\label{PSR}
Point sources in the FOV may contaminate the diffuse emission. They are identified and excluded using the XMM-SAS \texttt{cheese} script in this study.
We customize the \texttt{cheese} parameters to optimize point source detection.
First, the point spread function (PSF) threshold scale is fixed to 0.1, allowing the removal of a larger fraction of the point-source flux.
Second, the minimum separation for point sources is set to 2.5$\arcsec$, enabling the detection of point sources in crowded fields.
Finally, the point-source flux threshold rate is set to $3 \times 10^{-15}$ erg $\rm cm^{-2}$ $\rm s^{-1}$, improving the identification and exclusion of fainter sources. 
Following these adjustments, the \texttt{cheese} script is performed to search point sources in three energy bands: 400--1250 eV, 1250--7200 eV, and 400--7200 eV. This approach ensures the detection of soft X-ray sources, hard X-ray sources, and those with broad energy emissions. 

The script identifies the locations and sizes of detected point sources, recording them in a region file. 
For each observation, we combine the \texttt{cheese} detected regions from all EPIC-MOS exposures to create a master mask, leaving only the unmasked areas for subsequent analysis.
\blue{If the master mask covers $>50\%$ of the FOV, we exclude the observation to prevent potential contamination from extended sources in the FOV, which are not part of the SXRB we aim to measure.} As a result, 705 observations are removed because of this exclusion criterion.

\subsubsection{Extended X-ray Sources: Galaxy Clusters and Nearby Galaxies}
\label{EXS}

Galaxy clusters and nearby galaxies are extended sources emitting in the soft X-ray band. They serve as background contamination in studies of the MW hot gas or SWCX emission.
They can neither be excluded using \texttt{cheese} because they lack clear boundaries, nor be characterized by a spectral model due to their complex emission spectra.
To minimize this contamination, we estimate their angular size and mask them from the FOV. We assume all these sources are spherical for simplicity.

For galaxy clusters, we adopt the Meta-Catalogue of X-ray Detected Clusters of Galaxies (MCXC; \citealt{piffaretti2011mcxc}), a catalog of nearby X-ray bright galaxy clusters, which provides 
the redshift, physical radius ($r_{500}$\footnote{A characteristic radius defined as the distance from the center of a cluster to the location where the average mass density of the cluster drops to 500 times the critical density of the Universe.}), and position (i.e., Galactic longitude $l$ and latitude $b$). 
The angular size of each cluster is calculated using the $r_{500}$, beyond which the X-ray is negligible in observation.

For nearby galaxies, we adopt the \citet{kourkchi2017galaxy} galaxy catalog, which provides distance, position, and $K_s$-band luminosity over 15,000 galaxies within 3500 $\rm km$ $\rm s^{-1}$.
Since only a few massive galaxies have their hot gas halos detected \citep[e.g.,][]{Anderson:2010aa, Bogdan:2013aa}, we consider only those within 5 Mpc with stellar masses above $10^{10} ~\rm M_{\odot}$. Galaxies beyond 5 Mpc have detectable X-ray emissions primarily from their centers, which can be easily masked using the \texttt{cheese} script.
Stellar masses are derived using the $K_s$-band luminosity and the infrared mass-to-light ratio from \citet{kourkchi2017galaxy}. 
Besides the galactic center, the disk may also emit detectable X-rays \citep[e.g.,][]{li2013chandra}. Thus, we adopt an 8 kpc detection radius, a radius that covers most emissions from the nearest galaxies, such as the Triangulum Galaxy (M33), the Large Magellanic Cloud (LMC), and the Small Magellanic Cloud (SMC).

\begin{table*}[t!]
\caption{Classification of Bright X-Ray Observations}
\label{tab:flags}
\centering
\begin{tabular}{|l|c|l|}
\hline
\textbf{Category} & \textbf{Flag} & \textbf{Description} \\
\hline
\blue{Clean} & 0 & No detected contamination \\
\hline
\multirow{2}{*}{Extragalactic Objects} & 1 & Contaminated by galaxy clusters \\
\cline{2-3} 
& 2 & Contaminated by nearby galaxies \\
\hline
\multirow{5}{*}{Large-Scale MW Structures} & 3 & Projected within eROSITA bubbles \\
\cline{2-3} 
 & 4 & Cygnus Superbubble: centered at (80$\degr$, 2$\degr$) with axes of 17$\degr$ (long.) and 13$\degr$ (lat.) \\
\cline{2-3} 
 & 5 & Eridanus Superbubble: centered at (194$\degr$, -35$\degr$) with axes of 16$\degr$ (long.) and 17$\degr$ (lat.) \\
\cline{2-3} 
 & 6 & Vela SNR: centered at (261$\degr$, -3$\degr$) with axes of 9$\degr$ (long.) and 7$\degr$ (lat.) \\
\cline{2-3} 
 & 7 & Monogem SNR: centered at (197$\degr$, 10$\degr$) with radius of 16$\degr$\\
\hline
\multirow{2}{*}{Stellar Phenomena} & 8 & Other identified SNRs \\
\cline{2-3}
 & 9 & Bright stars \\
\hline
\multirow{2}{*}{Solar System Objects} & 10 & Comets \\
\cline{2-3} 
& 11 & Planets \\
\hline
\end{tabular}
\end{table*}

For each \xmm observation, we calculate the overlap area between the FOV and the extended sources using the circle-circle intersection formula: 
\begin{equation}
\begin{aligned}
A &=r^2 \arccos{\left(\frac{d^2 + r^2 - R^2}{2 d r}\right)}\\
&+R^2 \arccos{\left(\frac{d^2 + R^2 - r^2}{2 d R}\right)}\\
&-\frac{\sqrt{(-d+r+R)(d+r-R)(d-r+R)(d+r+R)}}{2},
\end{aligned}
\end{equation}
where $r$ is the FOV radius of EPIC-MOS fixed at $14\arcmin$, $R$ is the angular radius of the source, and $d$ is the angular separation between the source and the observation direction.

We exclude observations where the overlap area is $>$ 50\% of the FOV in the following analysis. For observations with $<$ 50\% overlap, we add their coordinates and estimated radii into the pre-generated master mask. This approach automatically excludes them along with those detected point sources. Figure \ref{fig:cheese} illustrates an example of the exclusion results for galaxy clusters and nearby galaxies.

This method may not always yield completely accurate results in all scenarios due to the assumptions involved in size estimation.
To address this, we introduce a ``flag" parameter. A flag value of ``1" or ``2" is assigned to observations in which galaxy clusters and nearby galaxies cover between 10\% and 50\% of the FOV (See Table \ref{tab:flags} for all observation flags). The ``flag'' parameter allows a manual exclusion of observations if needed.

In addition, an X-ray bright diffuse enhancement is discovered around M31, extending up to $\approx10\degr$--$20\degr$ \citep{qu2021x}, which is much larger than the estimated radius.
Thus, we manually flag all observations within 4.4 degrees from M31's center based on a visual inspection of the all-sky intensity maps as shown in Section \ref{sec:Results}. 

In total, 670 and 872 observations are excluded because of the large covering fractions ($>50\%$) of galaxy clusters and nearby galaxies, respectively.

\subsubsection{SXRB Spectra}

The SXRB spectra are extracted from the SP-filtered photon event lists using the entire FOV, excluding any anomalous CCDs and masked regions. This extraction is performed using the XMM-ESAS \texttt{mos-spectra} script, which also generates the corresponding redistribution matrix files (RMFs) and ancillary response files (ARFs).
Then, we use the \texttt{grppha} script in \texttt{FTOOLS} to group the SXRB spectra with associated QPB spectra, RMFs, and ARF files and bin them with a minimum of 100 counts per channel.
The SXRB spectra are extracted in the 0.3--3.2 keV band, which includes all three emission lines of interest (i.e., \ion{O}{7}, \ion{O}{8}, and Fe-L) and is broad enough to model the residual SPB.

\section{Components and Spectral Model} 
\label{sec:COMPONENTS AND SPECTRAL MODEL}

In the X-LEAP program, three spectral features (i.e., \ion{O}{7}, \ion{O}{8}, and Fe-L) are of particular interest to investigate the hot gas in the MW and the SWCX variations.
An accurate measure of their intensities requires precise spectral modeling to separate them from other components.
In this section, we describe the components considered in this work and their implementations in spectral fitting.

\begin{deluxetable*}{cccccc}
\scriptsize
\tablecaption{Spectral Model Settings in \texttt{XSPEC} \label{tab:MP}}
\tablewidth{0pt}
\tablehead{
\colhead{Component$\rm ^a$} & \colhead{Model$\rm ^b$} & \colhead{Parameter$\rm ^c$} & \colhead{Initial Value$\rm ^d$} & \colhead{Range$\rm ^e$}}
\startdata
Lines of Interest & \ion{O}{7} Gaussian &LineE (keV)& 0.565 & [0.535, 0.595] \\
& &Sigma (keV) & 0 & fixed\\
& & norm & 6.00 & $>$ 0\\
& \ion{O}{8} Gaussian &LineE (keV)& 0.654 & fixed\\
& &Sigma (keV) & 0 & fixed\\
& & norm & 2.00 & $>$ 0\\
& Fe-L Gaussian &LineE (keV)& 0.90 & [0.70, 1.00] \\
& &Sigma (keV) & 0.05 & fixed\\
& & norm & 0.80 & $>$ 0\\
\hline
MW CGM & APEC & kT (keV) & 0.20 & [0.10, 1.00] \\
& &Redshift & 0 & fixed\\
& & norm & 7.00 & $>$ 0\\
\hline
CXB & Powerlaw & PhoIndex & 1.46 & fixed\\
& & norm & 8.00 & [0, 50]\\
\hline
MW Absorption$\rm ^f$ & Phabs & nh & fixed & fixed\\
\hline
LHB & APEC & kT (keV) & 0.084 & fixed\\
& &Redshift & 0 & fixed\\
& & norm & 10.00 & $>$ 0\\
\hline
Residual SPB& Unfolded & PhoIndex & 0.90 & [0.10, 1.30] \\
& powerlaw &Redshift & 0 &fixed \\
& & norm & 10.00 & $>$ 0\\
\hline
Instrumental Lines& Al Gaussian &LineE (keV)& 1.49 & [1.46, 1.52] \\
& &Sigma (keV) & 0 & fixed\\
& & norm & 1.00 & $>$ 0\\
& Si Gaussian &LineE (keV)& 1.75 & [1.72, 1.78] \\
& &Sigma (keV) & 0 & fixed\\
& & norm & 1.00 & $>$ 0\\
\enddata
\tablenotetext{\rm{a}}{Emission sources considered in the SXRB.}
\tablenotetext{\rm{b}}{Model components associated with each emission source.}
\tablenotetext{\rm{c}}{Parameters required for each model component.}
\tablenotetext{\rm{d}}{Initial values assigned to these parameters.}
\tablenotetext{\rm{e}}{Allowed ranges for each parameter.}
\tablenotetext{\rm{f}}{Absorption model applied to account for dust and gas absorption attenuating MW CGM and CXB emissions. The column density for each observation is fixed to the total $N_{\rm H}$ derived from the 2018 \textit{Planck} dust emission survey (see Section \ref{subsec:gas absorption}).}
\end{deluxetable*}

\subsection{Overview of the Spectral Components}

\begin{figure*}[t]
\centering
\includegraphics[width=0.835\paperwidth]{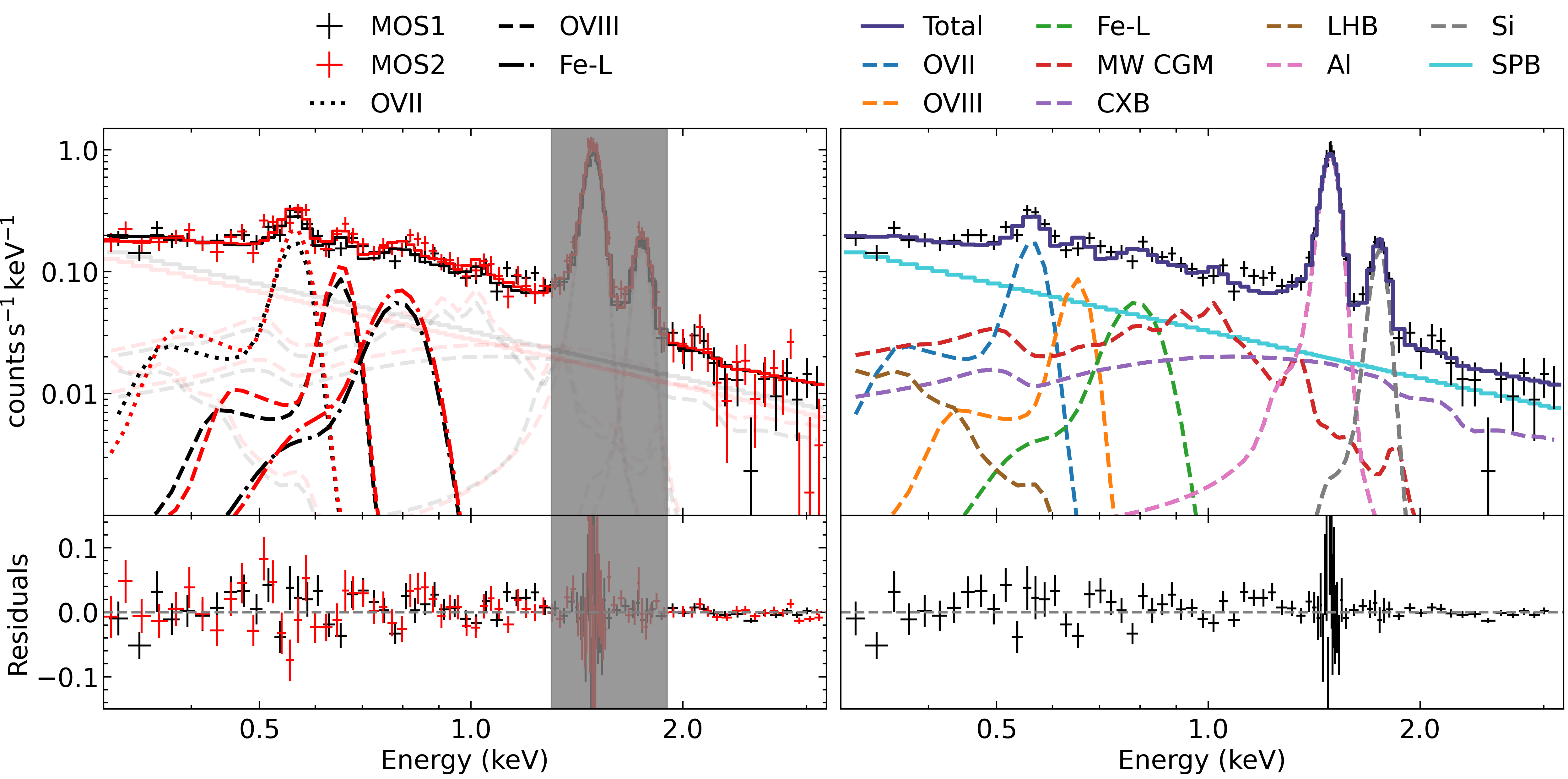}
\caption{An example of a joint-fitted SXRB spectrum and its best-fit model. Left: the paired spectrum extracted from ObsID: 0724770201. The \ion{O}{7} ($\approx$ 0.56 keV), \ion{O}{8} ($\approx$ 0.65 keV), and Fe-L ($\approx$ 0.80 keV) lines are modeled as three Gaussian functions and highlighted. Instrumental lines are covered by a grey shade. Right: MOS1 spectrum only. All spectral components are present illustrating individual contributions to the modeled spectrum. The soft proton background (SPB) is not folded through the instrumental response.}
\label{SpectralFitting}
\end{figure*}

The SXRB spectrum is composed of multiple spectral components, including the internal ``quiescent'' background (Section \ref{QPB}), the external ``flaring'' background (SPB; Section \ref{pep}), the SWCX emission, the LHB emission, the MW CGM emission, and the cosmic X-ray background (CXB) emission.
Both the X-ray emissions from the MW CGM and the CXB are subject to absorption by gas and dust in the MW (see Section \ref{subsec:gas absorption} for details).

The EPIC-MOS detectors are configured to observe the same sky regions simultaneously.
To amplify signals from diffuse hot gas emissions, we adopt a joint-fitting approach, treating both closely-timed EPIC-MOS spectra as a paired spectrum.
Specifically, we link all parameters corresponding to consistent components across both detectors in the fitting process.
Parameters associated with the instrumental backgrounds are not linked because the instrument background observed in one MOS detector slightly differs in the other \citep{snowden2014cookbook}.

The spectral fitting is done using \texttt{XSPEC} version 12.12.1\footnote{https://heasarc.gsfc.nasa.gov/xanadu/xspec/} \citep{arnaud1996xspec} with metal abundances from \cite{lodders2003solar}. This table is chosen for its widespread usage and because its oxygen and iron abundances align more closely with recent measurements, compared to the default table from \cite{anders1989abundances}. 
The following subsections detail how each component is modeled in \texttt{XSPEC}. The spectral components and their parameter settings are summarized in Table \ref{tab:MP}. 
In Figure \ref{SpectralFitting}, we show an example of a paired SXRB spectrum along with its best-fit model and all model components.

\subsection{The \ion{O}{7}, \ion{O}{8} and Fe-L Emission Lines}
\label{threelines}

\begin{figure*}[t!]
\centering
\includegraphics[width=0.83\paperwidth]{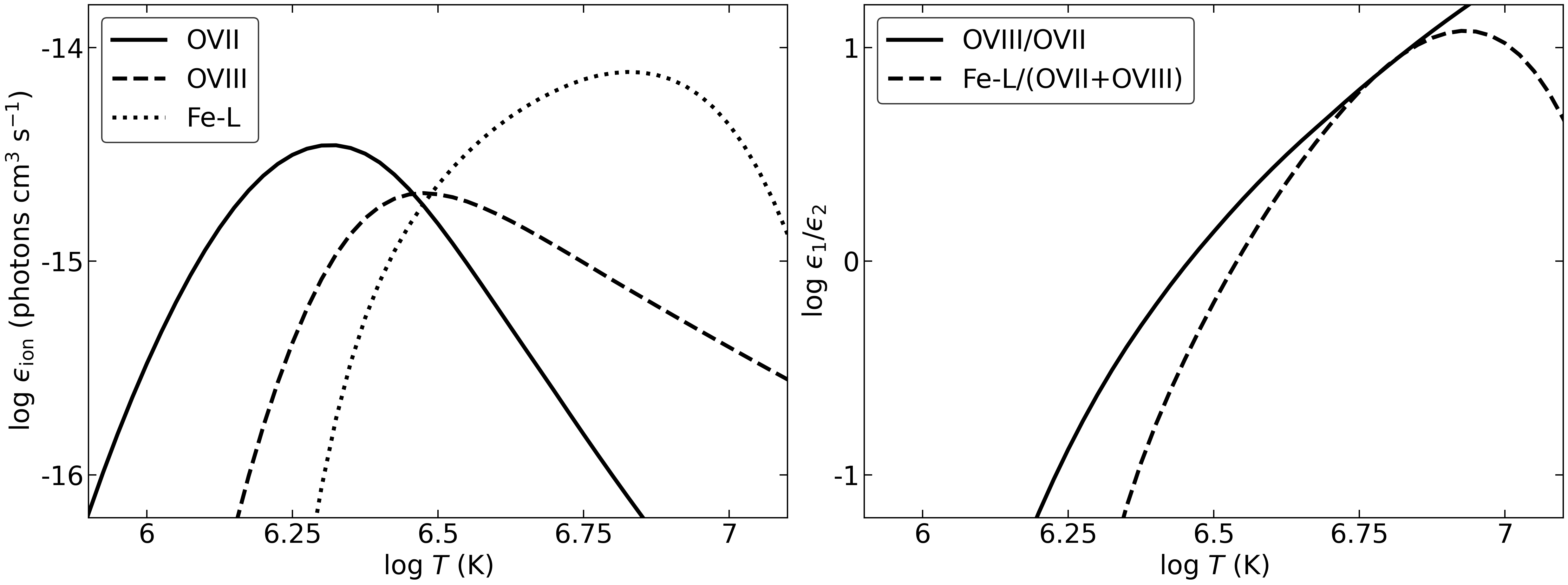}
\caption{Left: the \ion{O}{7}, \ion{O}{8}, and Fe-L line emissivities at different APEC temperatures \citep{foster2012updated}. \blue{They are disabled in the APEC model to prevent double-counting line intensities when extracting them using Gaussian functions.} Right: the emissivity ratios. The observed line intensity ratios are consistent with the emissivity ratios if the emission originates from a single-temperature plasma.}
\label{fig:emissivity}
\end{figure*}

The \ion{O}{7} K$\alpha$, \ion{O}{8} K$\alpha$, and Fe-L lines represent three spectral features in the SXRB. 
These lines do not refer to single ionic transitions.
Instead, each line is a set of transitions with similar transition energies originating from specific ions.

The \ion{O}{7} and \ion{O}{8} lines originate from both hot ionized gases and the SWCX process. 
The \ion{O}{7} line is a triplet consisting of the forbidden line at 561 eV, the intercombination line at 569 eV, and the resonance line at 574 eV. 
The \ion{O}{8} line is a doublet consisting of two close lines at 653.5 eV and 653.7 eV \citep{foster2012updated}.   
The EPIC-MOS spectral resolution is $\approx$ 40 eV at 500 eV \citep{lumb2012x}, much larger than the splitting of both the \ion{O}{7} triplet ($\approx$ 13 eV) and the \ion{O}{8} doublet ($<$ 1 eV), so we model these two lines as Gaussian functions.
The \ion{O}{7} line centroid is allowed to vary in the 535--595 eV energy range, whereas the \ion{O}{8} line centroid is fixed at 653.6 eV.
Given the spectral resolution exceeds the intrinsic line widths and the energy splitting, the observed line widths are mainly influenced by instrumental resolution. 
Consequently, we fix the line widths at 0, while leaving the normalizations of the Gaussian functions as free parameters.

The Fe-L line is composed of a set of iron emission lines in the $\approx 0.7-1.0$ keV range from \ion{Fe}{16} to \ion{Fe}{21}. The most prominent lines include \ion{Fe}{17} at $\approx$ 0.83 keV, \ion{Fe}{18} at $\approx$ 0.87 keV, and \ion{Fe}{19} at $\approx$ 0.92 keV \citep{foster2012updated}. These lines contribute the most to the total emissivity in the $\approx 0.7-1.0$ keV range at temperatures between $0.2 - 1.0 \times 10^{7}$ K. Previous studies suggest that this feature may be indicative of a hotter phase of the MW CGM with $kT \approx$ 0.41--0.72 keV \citep[e.g.,][]{das2019multiple, ponti2022abundance}.
However, it may be contaminated by the hot corona of unresolved stars \citep[e.g.,][]{wulf2019high}.
We investigate this feature to gain a deeper insight into the nature and origin of this feature.
It is modeled as a Gaussian function with the line centroid varying within the 0.7--1.0 keV energy range. The line width is fixed at 50 eV, consistent with the standard deviation of the transition energies of these iron lines. The normalization is varied as a free parameter. 

We employ the Astrophysical Plasma Emission Code (APEC) model for MW and LHB emissions \citep{smith2001collisional}, which will be introduced in the following sections. To avoid double-counting the line emissions, we disable them in the APEC model. 
Adopting the approach of \cite{lei2009determining}, we set emissivities of the \ion{O}{7} and \ion{O}{8} K$\alpha$ lines to zero in the APEC line emissivity file (apec\_v3.0.9\_line.fits, \citealt{foster2012updated}). Note that we do not disable the \ion{O}{7} K$\beta$ line at 666 eV near the \ion{O}{8} K$\alpha$ line at 653 eV. This may slightly lower the observed \ion{O}{8} intensity.
For the Fe-L emission, we examine the emissivity data in the same file within $kT=0.5$--0.8 keV. At each temperature, we identify the 50 strongest emission lines and disable only iron lines by setting their emissivities to zero. 

Figure \ref{fig:emissivity} shows the \ion{O}{7}, \ion{O}{8}, and Fe-L emissivities at different APEC temperatures using abundances from \cite{lodders2003solar}. It demonstrates that in a single-temperature plasma under CIE, the emissivities for \ion{O}{7}, \ion{O}{8}, and Fe-L emissivities peak at $\approx2\times10^6$ K, $\approx3\times10^6$ K, and $\approx7\times10^6$ K, respectively. Since line intensity scales as $I \propto n_e^{2} \epsilon (T) Z$, a high intensity implies a plasma temperature close to these peak values. Moreover, the intensity ratios, such as $I_{\rm O VII}/I_{\rm O VIII} \propto \epsilon_{\rm O VII} (T)/ \epsilon_{\rm O VIII} (T)$, provide a direct measure of plasma temperature. These temperature diagnostics have been extensively explored in \citet[][hereafter \citetalias{Qu:2023aa}]{Qu:2023aa}.

\subsection{the MW CGM}

The diffuse gas surrounding the MW, known as the MW CGM, contains a hot phase with temperatures typically ranging from 1--$3 \times 10^6$ K, leading to X-ray emissions. Such hot gas is thought to be an optically thin plasma in CIE and is commonly modeled using the APEC model (represented as \texttt{apec} in \texttt{XSPEC}; \citealt{smith2001collisional}).  
In our MW APEC model, we fix the metal abundance at 0.5 $Z_\odot$ but allow temperature and normalization to vary. \blue{The hot gas is expected to have temperatures $kT \approx 0.1$--0.3 keV, which is most sensitive to the \ion{O}{7} and \ion{O}{8} lines we disabled \citep{miller2016interaction}.
Thus, we adopt a wider allowed temperature range of 0.1--1.0 keV for this model.}

It is worth noting that a cooler phase with $kT < 0.1$ keV might exist \citep[e.g.,][]{Kuntz:2000}, potentially enhancing the lower end of the SXRB spectrum. However, this component is barely detectable by \xmm or \textit{Chandra} and would likely overlap with the LHB component described in Section \ref{subsec:LHB}. Furthermore, our line extraction method directly measures line intensity from the spectrum, making it insensitive to assumptions regarding hot gas models, as discussed in Introduction \ref{sec:intro}.

\subsection{The CXB}

The CXB is defined as the total emission from all the extragalactic sources in the X-ray band, which dominates the SXRB spectrum above 1 keV. 
The primary contributors to the CXB emission are active galactic nuclei (AGNs), with some contributions from galaxy clusters and starburst galaxies \citep{Gilli:2007aa, moretti2009new}.
The spectral characteristics of the CXB, largely influenced by the AGNs, typically exhibit power law distributions. This is a consequence of synchrotron radiation, a common emission mechanism in AGNs \citep{piconcelli2005xmm}.
We model the CXB using a power-law (\texttt{powerlaw} in \texttt{XSPEC}) with a photon index fixed to $\Gamma=1.46$, which represents the average spectral shape of the distant AGN population \citep{chen1997asca}. 
Although the CXB is generally isotropic, it still exhibits some level of variation across the sky. Therefore, the normalization of CXB remains a free parameter for each observation.

\subsection{MW Absorption}
\label{subsec:gas absorption}
Emissions from both MW CGM and CXB travel through cold dust and clouds, leading to absorption in soft X-rays.
This absorption is modeled using the \texttt{phabs} absorption model \citep{balucinska1992photoelectric} in \texttt{XSPEC}, with an updated He cross-sections from \citet{yan1998photoionization}.
The absorption efficiency is determined by the hydrogen column density ($N_{\rm H}$), a key input for the \texttt{phabs} model.

We estimate $N_{\rm H}$ values based on the 2018 \textit{Planck} dust emission survey, which measures the dust opacity at 353 GHz ($\tau_{353}$;  \citealt{Planck:2014}). This opacity traces both the molecular and atomic portions of the absorbing medium, a good tracer for $N_{\rm H}$ particularly in regions containing both $\rm H_2$ and \ion{H}{1}. Specifically, the dust-inferred $N_{\rm H}$, $N_{\rm H, dust}$, is calculated based on the visible extinction map\footnote{https://pla.esac.esa.int/\#maps} derived from the dust survey, using the relation $N_{\rm H}\ (\rm 
cm^{-2})= 2.21 \times 10^{21}$ $A_{\rm V}$ \citep{Guver:2009}.
For comparison, we also calculate the \ion{H}{1}-inferred $N_{\rm H}$ using the HI4PI survey \citep{bekhti2016hi4pi}.

For each observation, we determine the $N_{\rm H}$ using the averaged $N_{\rm H}$ within a 0.3$\degr$ radius, slightly larger than the FOV of EPIC-MOS.

The results show that the $N_{\rm H, dust}$ values align with those inferred from \ion{H}{1} in over 50\% of the observations, specifically at $|b|>30\degr$, without exceeding by more than a factor of one. However, at $|b|<30\degr$ --- a region rich in molecular hydrogen --- the $N_{\rm H, dust}$ median is approximately 2.5 times higher than the \ion{H}{1} inferred $N_{\rm H}$ median. This significant difference suggests that $N_{\rm H, dust}$ more accurately accounts for the total absorbing medium, making it a better tracer, specifically in $\rm H_2$ dense regions.

The derived $N_{\rm H, dust}$ values are also used in Section \ref{sec:Soft X-ray distribution of the MW} for estimating the deabsorbed MW emission.

\subsection{The LHB}
\label{subsec:LHB}
The LHB is a cavity filled with hot and diffuse gas surrounding the Solar System. It extends $\approx 100$ pc in all directions away from the Sun \citep{liu2016structure}, with an estimated temperature of $kT \approx 0.084$ keV (e.g., \citealt{bluem2022widespread}; \citealt{yeung2023srg}) and a relatively constant electron density of $4 \times 10^{-3}$ $\rm cm^{-3}$ \citep{yeung2023srg}.
We model the LHB's emission using another APEC model with the temperature, abundance, and redshift parameters fixed at 0.084 keV, 1.0, and 0, respectively. The normalization remains a free parameter.
We note that the \ion{O}{7}, \ion{O}{8}, and Fe-L lines are disabled in both the MW CGM and LHB APEC models \blue{to prevent double-counting these intensities when measuring them directly using Gaussian functions.}
\subsection{Residual Soft Proton Contamination}
\label{subsec:Residual SP cotamination}
The SPB is part of the instrumental background, which primarily comes from soft protons interacting directly with the detector. It can not be fully removed by the flare-filtering process discussed in Section \ref{pep}.
Here, we model the residual SP contamination using a power-law not folded through the instrumental effective areas, since these protons do not interact with the parts of the instrument that shape the effective area. 
The power-law index parameter is allowed to vary between 0.2--1.3 following the XMM-ESAS manual \citep{snowden2011cookbook}.
The normalization is a free parameter, allowed to vary across observations.

\subsection{The Instrumental Lines}
\label{subsec:InstrumentalLines}
As previously discussed in Section \ref{QPB}, the Al and Si instrumental lines at 1.49 keV and 1.74 keV can not be adequately characterized by the \texttt{mos\_back} script. 
They are modeled as two Gaussian functions with line centroids confined to the ranges of 1.46--1.52 keV and 1.72--1.78 keV, respectively. 
The line widths are fixed at 0, and the normalizations remain as free parameters.

\begin{deluxetable*}{ccccccccccc}
\centering
\tablecaption{Properties of Observed \ion{O}{7}, \ion{O}{8}, and Fe-L Lines$\rm^a$ \label{tab:LT}}
\tablewidth{0pt}
\tablehead{
\colhead{ObsID}                       & \colhead{Pair$\rm^b$}                       &
\colhead{Date}                      & \colhead{GTI}                   &
\colhead{$l$}                        & \colhead{$b$}                        &
\colhead{$E_{\rm OVII}$$\rm^c$}             & \colhead{$I_{\rm OVII}$$\rm^d$}  &
\colhead{$I_{\rm OVIII}$$\rm^d$} & \colhead{$I_{\rm FeL}$$\rm^d$}   &
\colhead{Flag$\rm^e$}                       \\[-2ex] 
\colhead{}           & \colhead{}          &
\colhead{}           & \colhead{(ks)}      &
\colhead{(deg)}      & \colhead{(deg)}     &
\colhead{(eV)}      & \colhead{(L.U.)}    &
\colhead{(L.U.)}     & \colhead{(L.U.)}    &
\colhead{}           \\[-1ex]
\colhead{(1)}  & \colhead{(2)}  &
\colhead{(3)}  & \colhead{(4)}  &
\colhead{(5)}  & \colhead{(6)}  &
\colhead{(7)}  & \colhead{(8)}  &
\colhead{(9)}  & \colhead{(10)} &
\colhead{(11)}}
\startdata
0606430401 & 1S001, 2S002 & 2009-12-07 &      15.7 &  18.77 & -84.80 &   $557.0_{-5.7}^{+6.0}$ &  $3.78_{-0.66}^{+0.69}$ &  $0.92_{-0.24}^{+0.29}$ &  $1.18_{-0.32}^{+0.32}$ &    0 \\
0723180601 & 1S001, 2S002 & 2014-03-07 &       8.0 & 143.78 &  19.40 & $555.1_{-17.5}^{+23.4}$ &                $<$ 3.79 &  $1.64_{-0.78}^{+0.57}$ &  $2.89_{-0.43}^{+0.53}$ &    0 \\
0204190201 & 1S001, 2S002 & 2004-02-09 &       5.0 & 225.82 & -56.31 & $559.6_{-14.8}^{+17.9}$ &  $5.53_{-1.59}^{+1.94}$ &                $<$ 3.26 &  $0.94_{-0.51}^{+0.40}$ &    0 \\
0111282001 & 1S001, 2S002 & 2002-06-22 &       7.9 & 234.16 & -88.68 &   $574.8_{-5.5}^{+5.9}$ &  $6.03_{-0.95}^{+0.82}$ &                $<$ 1.27 &                $<$ 1.07 &    0 \\
0605391101 & 1S001, 2S002 & 2009-12-26 &       7.4 & 138.53 & -62.07 & $556.7_{-15.3}^{+13.1}$ &  $2.04_{-1.06}^{+0.83}$ &                $<$ 1.43 &                $<$ 1.99 &    0 \\
0201901701 & 1U002, 2U002 & 2004-07-28 &       7.8 & 299.45 &  28.87 &   $558.0_{-6.4}^{+4.9}$ & $15.89_{-1.87}^{+2.96}$ &  $3.62_{-0.66}^{+0.69}$ &  $2.38_{-0.75}^{+0.81}$ &    1 \\
0201902701 & 1S001, 2S002 & 2004-10-28 &      17.1 &  33.32 & -48.44 &   $562.1_{-4.9}^{+2.8}$ & $12.12_{-1.06}^{+0.81}$ &  $3.67_{-0.51}^{+0.95}$ &                 $<$ 1.8 &  1,3 \\
0673580301 & 1S001, 2S002 & 2011-07-25 &      74.6 & 310.19 &  23.98 &   $562.3_{-1.3}^{+2.2}$ & $25.85_{-0.55}^{+0.69}$ & $12.72_{-0.33}^{+0.28}$ & $10.22_{-0.21}^{+0.32}$ &    2 \\
0721900101 & 1S007, 2S009 & 2013-05-03 &      11.2 & 220.24 &  21.97 &   $576.2_{-7.0}^{+5.6}$ &  $4.74_{-0.73}^{+1.04}$ &                $<$ 1.51 &                $<$ 1.78 &    0 \\
0721900101 & 1U002, 2U002 & 2013-05-03 &      25.6 & 220.24 &  21.97 &   $561.8_{-3.6}^{+4.5}$ &  $5.35_{-0.50}^{+0.62}$ &  $1.15_{-0.37}^{+0.36}$ &  $1.01_{-0.45}^{+0.37}$ &    0 \\
\multicolumn{1}{c}{\ldots} & \multicolumn{1}{c}{\ldots} & \multicolumn{1}{c}{\ldots} & \multicolumn{1}{c}{\ldots} & \multicolumn{1}{c}{\ldots} & \multicolumn{1}{c}{\ldots} & \multicolumn{1}{c}{\ldots} & \multicolumn{1}{c}{\ldots} & \multicolumn{1}{c}{\ldots} & \multicolumn{1}{c}{\ldots} & \multicolumn{1}{c}{\ldots} 
\enddata
\tablenotetext{\rm{a}}{Ten typical samples from the catalog. The complete dataset is available in a machine-readable table online.}
\tablenotetext{\rm{b}}{Paired EPIC-MOS exposures observe the same sky region simultaneously. If an observation has more than one pair, they are listed separately, as illustrated by the last two rows in the table.}

\tablenotetext{\rm{c}}{The median and 68\% credible interval for the \ion{O}{7} line centroid, derived from MCMC chains.}
\tablenotetext{\rm{d}}{Intensities for \ion{O}{7} and \ion{O}{8} K$\alpha$, and Fe-L spectral lines extracted from MCMC chains. Medians and 68\% credible intervals are reported for measurements, while the 95\% percentile is reported for upper limits.}
\tablenotetext{\rm{e}}{Classification of bright X-ray observations for manual data filtering.}
\end{deluxetable*}

\subsection{Model Fitting}
\label{subsec: Model Fitting}
In \texttt{XSPEC}, our spectral model is structured as \texttt{constant(phabs(apec$_{\text{MW}}$ + powerlaw$_{\text{CXB}}$) + gaussian$_{\text{OVII}}$ + gaussian$_{\text{OVIII}}$ + gaussian$_{\text{FeL}}$ + apec$_{\text{LHB}}$) + gaussian$_{\text{Al}}$ + gaussian$_{\text{Si}}$}, complemented by a residual SPB (\texttt{powerlaw$_{\text{SPB}}$}) unfolded with the response matrix.
From left to right, the \texttt{constant} accounts for the total usable FOV for the detector, which is calculated by the \texttt{scale\_proton} script in \texttt{XSPEC}.
This factor allows us to directly derive FOV averaged parameters. 
The \texttt{phabs(apec$_{\text{MW}}$ + powerlaw$_{\text{CXB}}$)} shows how the absorption of MW hot gas emission and the CXB is corrected in \texttt{XSPEC}.
The following three Gaussian functions represent the three lines of interest.
The \texttt{apec$_{\text{LHB}}$} after that represents the unabsorbed LHB emission.
The two Gaussian functions at the end represent the Al and Si instrumental lines.

The model fitting is performed in two steps. 
We first use the $\chi^2$-statistic in \texttt{XSPEC} to fit each paired spectrum, providing initial estimates for the best-fit parameters. 
Although $\chi^2$ fitting provides a quick and reasonable parameter estimation, it can sometimes find local minima and yield less accurate uncertainty estimations, particularly with multiple free parameters. 
Therefore, we adopt the Markov Chain Monte Carlo (MCMC) method secondly after the $\chi^2$ fitting. This approach allows for a more robust and efficient exploration of the parameter space. 

The MCMC is performed with the Goodman-Weare algorithm \citep{goodman2010ensemble} via the \texttt{chain type} command in \texttt{XSPEC}. 
\blue{It starts with the best-fit parameters from the $\chi^2$ fitting as initial values for the walkers, using the \texttt{chain gaussian deltas 100} command.}
Considering 20 free parameters in the model, 40 MCMC walkers are used with each run for a minimum of 10,000 steps. Convergence is continuously monitored using the Gelman-Rubin convergence diagnostic \citep[$R$, ][]{Gelman:1992}, a widely used method to assess MCMC chain convergence. A value of $R<1.1$ generally indicates good convergence. If any spectral line parameters fail to meet this threshold, the fitting process is extended with additional steps until convergence is achieved. Our data analysis uses only the last 2000 points from these MCMC chains.

In some cases, the signal from the hot gas and SWCX may be weak or indistinguishable from noise. To distinguish real signals from noise, we apply the following criteria: if the median of an MCMC chain exceeds the width of its 68\% credible interval, we classify it as a measurement and report the median and the 68\% credible interval. If not, we classify it as an upper limit and only report the 95th percentile, indicating that we are 95\% certain the intensity is below this value.

\subsection{Visual Inspection of Fitting Results}
\label{subsec: Visual Inspection}
Although the spectral model is specifically designed for SXRB spectra, it may not fit with cases where bright X-ray sources are not fully masked or excessive SP contamination remains during high SPB activity. These could bias line intensity measurements, necessitating manual exclusion of such observations through visual inspection.
After inspecting each spectrum (e.g., Figure \ref{SpectralFitting}) and its best-fit model, we identify dozens of poorly-fitted spectra caused by the reasons above and resulting in significant residuals at certain energies. Additionally, a few spectra show a feature approximately 10 eV lower than the median energy of the \ion{O}{7} line centroids. \red{Such unexpected diffuse emissions are likely shifted \ion{O}{7} emissions from unmasked diffuse sources beyond the MW.}
Therefore, we exclude observations with $\chi^2/\rm{DoF}>5$ or $E_{\rm OVII}$ less than 555 eV, considering the spectral resolution and the instrumental gain (see Section \ref{subsec:O7 centroid} for more detail).

The remaining problematic observations are concentrated in low galactic latitude regions.
These regions, dense with gas and dust, exhibit strong X-ray absorption and complex emissions from the Galactic disk, complicating line measurements.
Therefore, we exclude observations within $|b|<2\degr$, the region where most problematic observations are found.

Based on visual inspection and low latitude exclusion, we further exclude 1921 observations from our sample. This results in a final sample of 5418 observations or 5470 paired exposures.

\section{Measurement Results}
\label{sec:Results}

\subsection{Catalog}
\label{sub:Catalog}
The line measure results are presented in Table \ref{tab:LT}, along with corresponding observational information.
Columns in this table include (1) the \xmm observation ID; (2) the name of each EPIC-MOS pair; (3) the observation start date in YYYY-MM-DD; (4) the mean GTI of each EPIC-MOS pair; (5) and (6) the observation direction in galactic coordinates $(l, b)$; (7) the \ion{O}{7} line centroid ($E_{\rm O VII}$) median with 68$\%$ credible interval; (8), (9), and (10) the observed \ion{O}{7}, \ion{O}{8}, and Fe-L intensities. Measurements are reported with medians and 68\% credible intervals, while upper limits are given at the 95th percentile. \blue{The criteria for distinguishing these two are in Section \ref{subsec: Model Fitting}}; and (11) the classification flag (detailed in Section \ref{OF}). The observed intensities of the three lines are illustrated in all-sky maps shown in Figure \ref{fig:observedAllsky}.

\subsection{Classification of Bright Observations}
\label{OF}

\begin{figure}[ht!]
    \centering
    
    \begin{minipage}[ht!]{0.475\textwidth}
        \includegraphics[width=\linewidth]{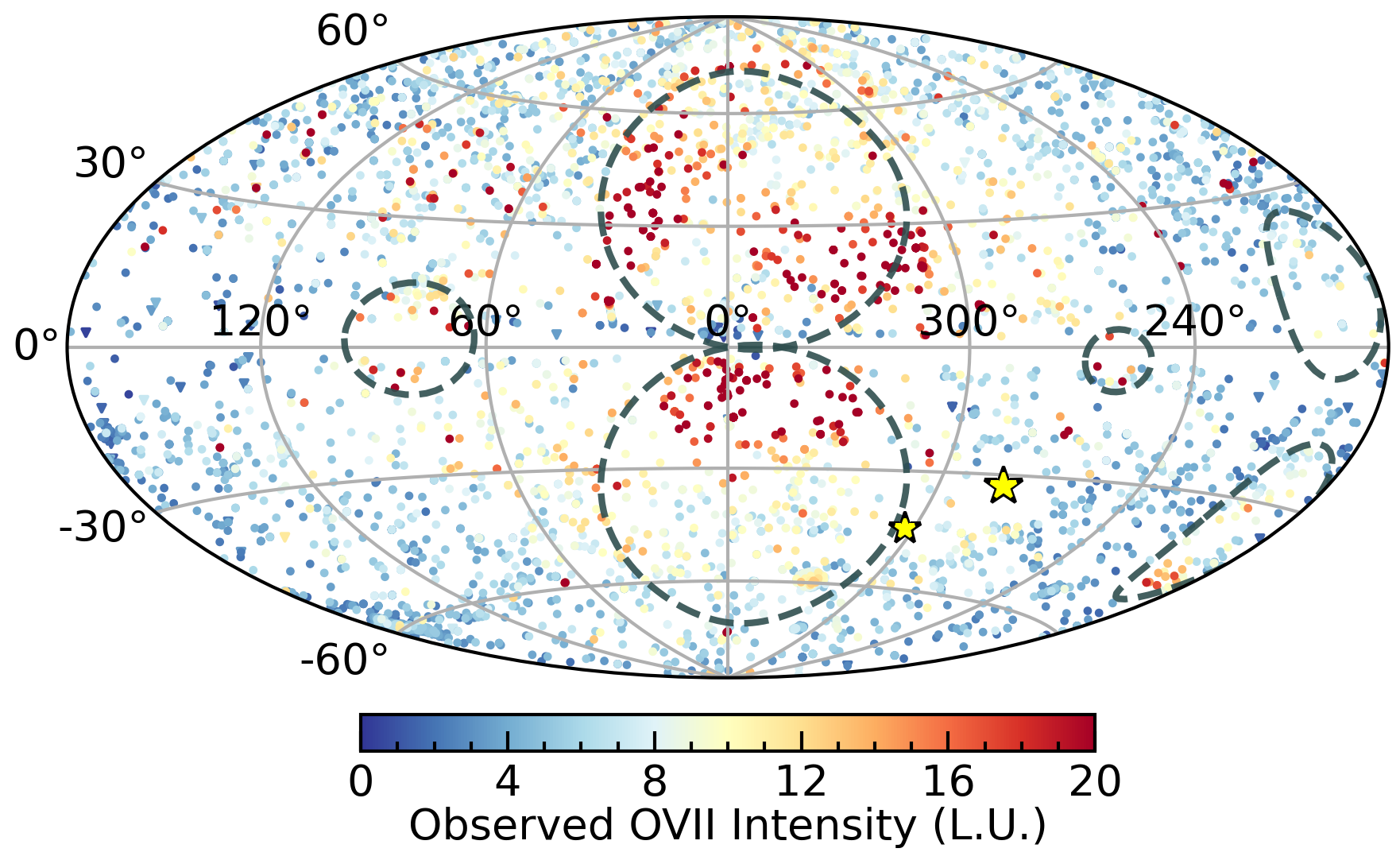}
    \end{minipage}
    \hfill
    \begin{minipage}[ht!]{0.475\textwidth}
        \includegraphics[width=\linewidth]{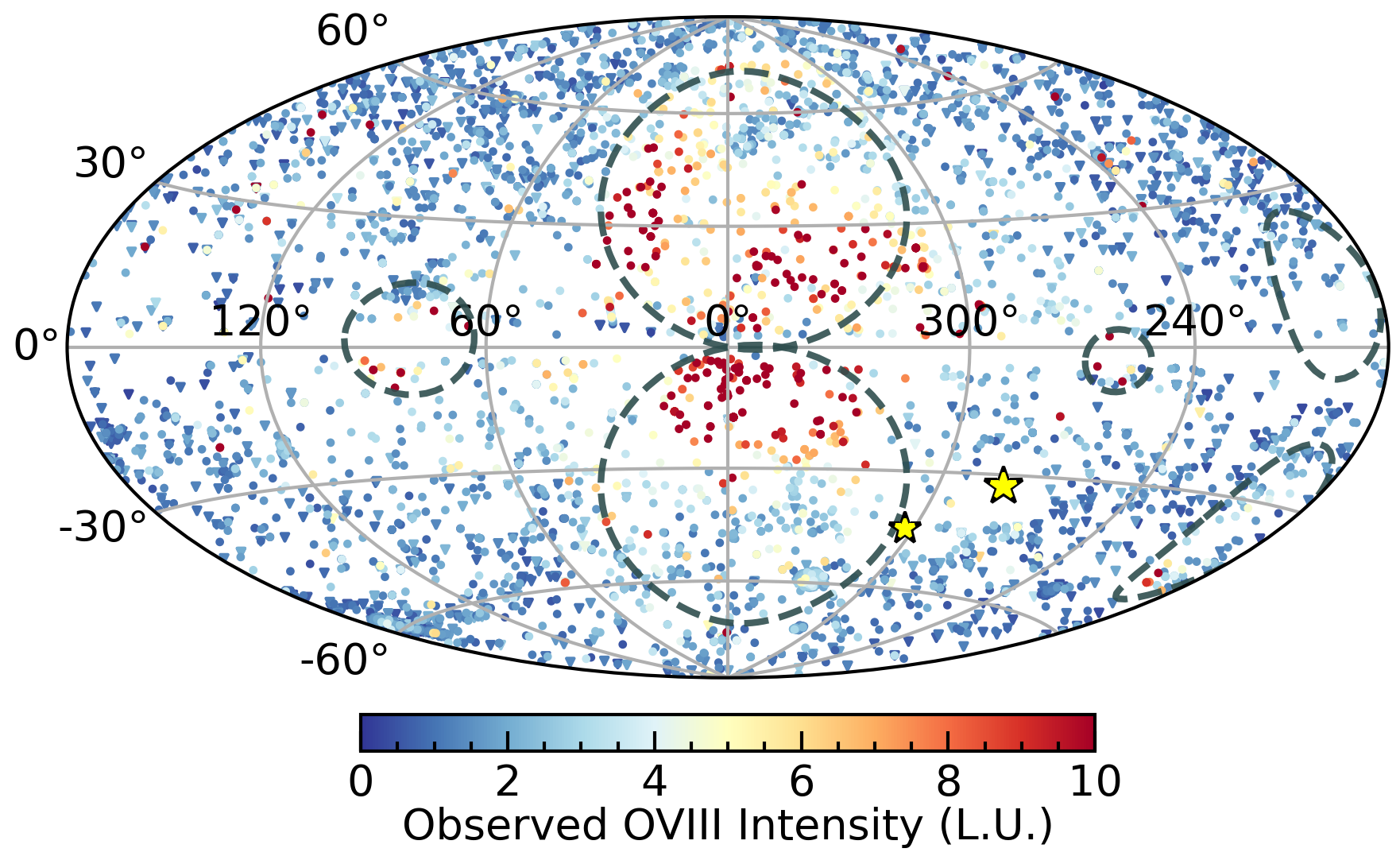}
    \end{minipage}
    
    \begin{minipage}[ht!]{0.475\textwidth}
        \includegraphics[width=\linewidth]{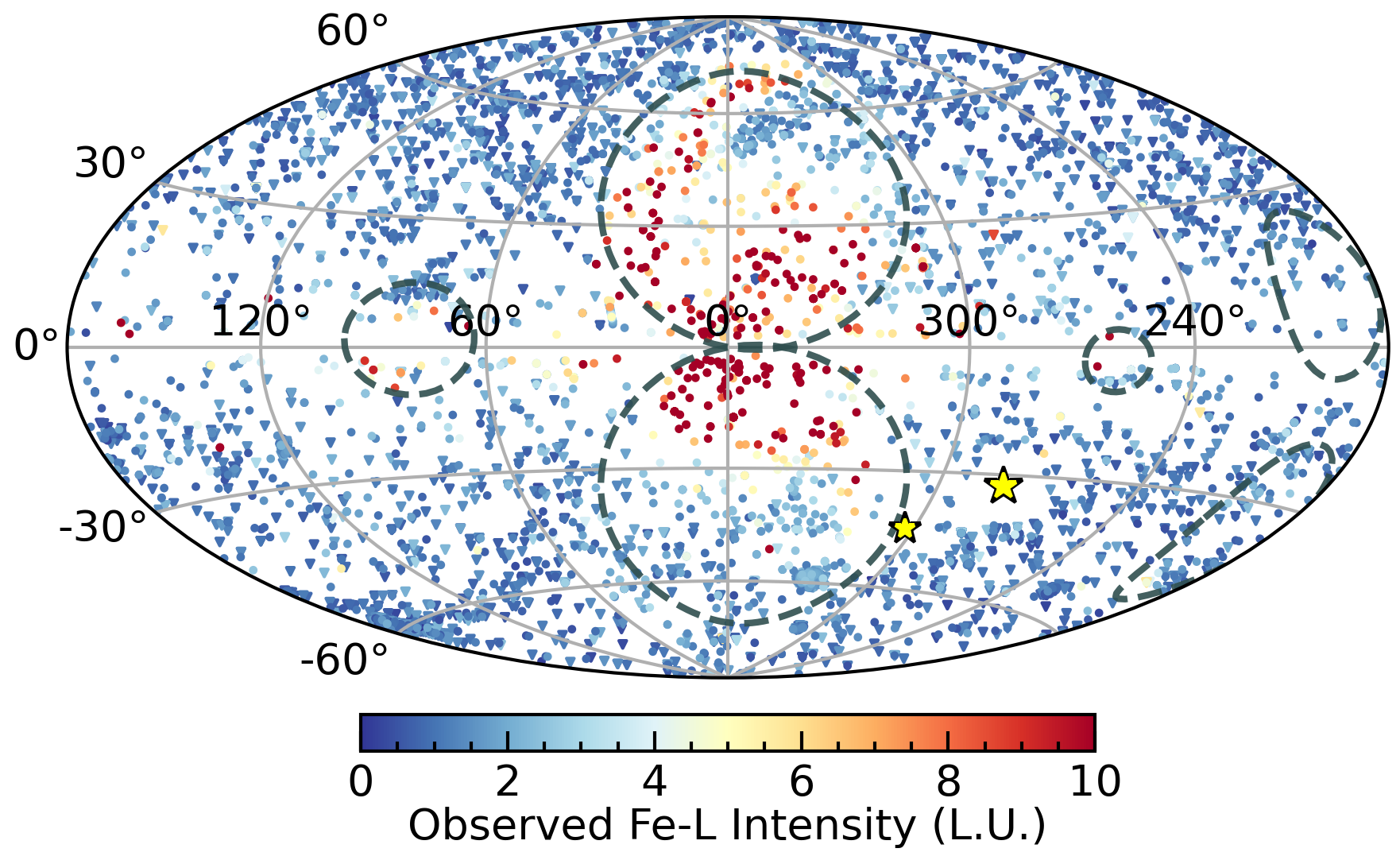}
    \end{minipage}
    
\caption{From top to bottom: all-sky maps of observed \ion{O}{7}, \ion{O}{8}, and Fe-L intensities, shown in Hammer projections centered on the Galactic center. These observations cover approximately 3\% of the sky. Measurements (dots) and upper limits (triangles) are plotted according to their pointing directions. Dashed lines enclose five large MW structures: the Cygnus Superbubble ($80\degr$, $2\degr$), the eROSITA bubbles (centered at Galactic center), the Vela and Monogem supernova remnants (($261\degr$, $-3\degr$) and ($197\degr$, $10\degr$)), and the Eridanus Superbubble ($194\degr$, $-35\degr$). The Large and Small Magellanic Clouds are marked with star symbols.}
\vspace{4ex}
\label{fig:observedAllsky}
\end{figure}

The SXRB measurements can be enhanced by various astronomical objects, including SWCX around comets or solar planets within the Solar System, Galactic bubbles, diffuse superbubbles (SBs) or supernova remnants (SNRs) within the MW, and galaxy clusters or galaxies beyond the MW.
They may not exhibit clear structures within the FOV, yet can enhance the intensity measurements.
Their contamination may not be fully excluded using the procedures in Section \ref{sec:DataReduction}, leading to the need for additional data filtering.

\begin{figure*}[ht!]
\centering
\includegraphics[width=0.8\paperwidth]{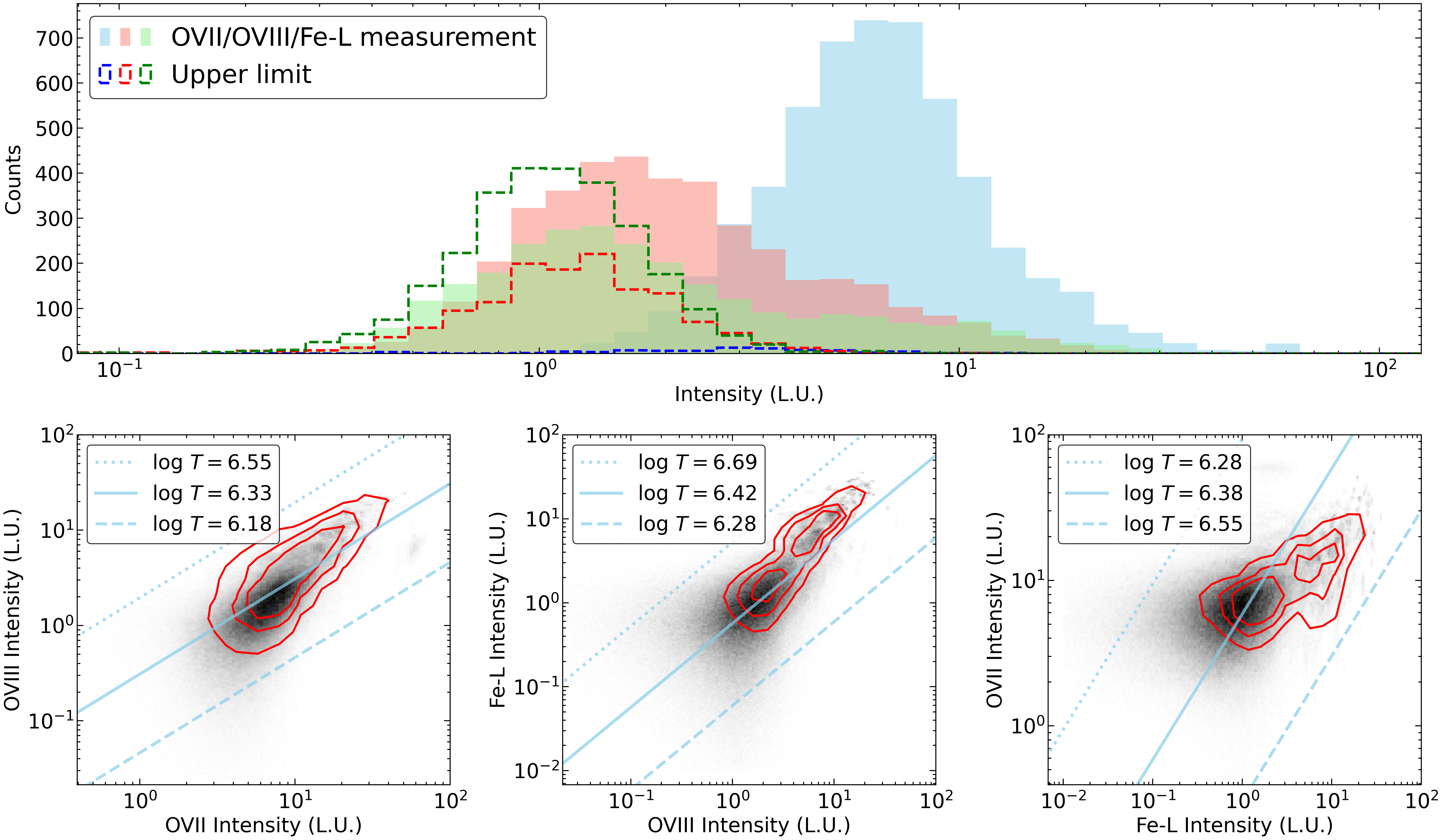}
\caption{Top: intensity distributions of \ion{O}{7}, \ion{O}{8}, and Fe-L lines. Measurements and upper limits are shown in bar-type and dashed step-type histograms, respectively. Bottom: correlations of observed \ion{O}{7}, \ion{O}{8}, and Fe-L intensities. Each panel displays the two-dimensional histograms of intensities (grayscale) using points from the MCMC chain. The blue dashed lines represent the CIE model predictions at different temperatures, while the contours (red) represent intensities measured inside eROSITA bubbles (eRBs). The eRBs contribute most of the high-intensity emissions, with some exceptions due to small X-ray structures or X-ray bright phenomena outside the eRBs.
Positive relations between emission lines imply a consistent temperature of the MW's gas, with a relatively narrow temperature dispersion. A detailed study on MW temperature structures is in \citetalias{Qu:2023aa}.}
\label{fig:I_dist}
\end{figure*}

We add a classification parameter, ``Flag", in our catalog to classify observations, which are likely contaminated by these objects. The classification is based on the presence of X-ray bright objects in the FOV or the observation direction. 
Observations contaminated by known galaxy clusters and nearby galaxies are flagged with ``1" and ``2" in previous Section \ref{EXS}.
We further expand the initial classification by visually inspecting the all-sky maps in Figure \ref{fig:observedAllsky}. We focus on isolated observations with measured line intensities at least three times higher than their neighboring observations. These
observations are then classified based on their target names and/or observation directions. 

In addition, we notice five extended regions with local intensity enhancements: eROSITA bubbles (eRBs; \citealt{predehl2020detection}), Cygnus SB \citep{cash1980x}, Eridanus SB \citep{ochsendorf2015nested}, Vela SNR \citep{helfand2001vela}, and Monogem SNR \citep{thorsett2003pulsar}.
They are remnants of past astronomical events with angular radii $\sim10\degr$. Observations within these regions are directly flagged based on their angular sizes in soft X-rays.
The size of eRBs is approximated by projecting ellipsoids at the Galactic center with dimensions of $14 ~\rm kpc \times 9 ~kpc \times 9$ kpc and tilted by 10$^\circ$ to align with observations. The sizes of the other four are estimated based on visual inspection of the line intensity maps in Figure \ref{fig:observedAllsky}. All classification flags are summarized in Table \ref{tab:flags}. 

We note that the clean sample (Flag=``0") is most suitable for diffuse gas studies due to the absence of contamination from bright X-ray sources. Observations flagged ``1" and ``2" are generally usable, as we mitigate contamination from galaxy clusters and nearby galaxies (see Section \ref{EXS}), though a few may still have notable contamination from extended sources. Flags of ``3'' to ``7'' indicate observations within five large X-ray structures. They are high-intensity regions resulting from past events and not fully mixed with the hot CGM. Observations with Flags of  ``8'' to ``11'' should only be used as references for other studies as they do not represent the MW's hot gas emission.

\subsection{Line Intensity Properties}
\label{sec:Line intensity properties}

The observed \ion{O}{7}, \ion{O}{8}, and Fe-L line intensities have medians of $\approx$ 6 photons cm$^{-2}$ s$^{-1}$ sr$^{-1}$ (line unit; L.U.), $\approx$ 1 L.U., and $\approx$ 1 L.U., respectively. Furthermore, 90\% of these intensities fall within the ranges of $\approx$ 2--18 L.U., $\approx$ 0--8 L.U., and $\approx$ 0--9 L.U.
Here, we explore the intensity distributions and correlations between these lines and discuss the implications of these relationships.
Figure \ref{fig:I_dist} presents two-dimensional histograms, showing the correlations between the intensities of these emission lines. Each observation is represented by all points from the MCMC chain.
Blue dashed lines represent the collisional ionization equilibrium (CIE) model predictions at different temperatures. 

All panels show strong positive correlations, indicating a roughly consistent intensity ratio. This consistency, as predicted by the CIE model, points to a relatively narrow temperature range in MW hot gas.
A more detailed study on MW's temperature distribution is presented in \citetalias{Qu:2023aa}.
In addition, the red contours enclosing the eRBs' measurements indicate that the eRBs contribute most of the high-intensity measurements. The double peak in the contours suggests that the eRBs may consist of two components, which may be evidence of multi-phase structures within the eRBs.
Furthermore, the presence of high-intensity observations outside the eRBs suggests the existence of other high-intensity emission regions in the hot gas. These regions may represent other small X-ray structures or X-ray bright phenomena within the hot gas.

\subsection{Data Comparison}
\label{Data Comparison}
In this section, we compare our measurements of \ion{O}{7} and \ion{O}{8} to those reported by \citetalias{henley2012xmm}. 
There are several differences in our methodology from that of \citetalias{henley2012xmm}, which could potentially result in inconsistent \ion{O}{7} and \ion{O}{8} measurements:

\begin{enumerate}
\item We use version 20.0.0 of the SAS software, whereas \citetalias{henley2012xmm} use version 11.0.1.
\item We use \texttt{cheese} to mask out point sources and two catalogs to mask out known galaxy clusters and nearby galaxies. In contrast, \citetalias{henley2012xmm} use the \xmm Serendipitous Source Catalogue to mask out point sources and some other sources by hand.
\item The metal abundances in our study are from \cite{lodders2003solar}, while the metal abundances use in \citetalias{henley2012xmm} are from \cite{anders1989abundances}.
\item We treat the CXB normalization parameter as a free parameter in our model, while \citetalias{henley2012xmm} fixes it at 7.9 $\mathrm{cm^{-2}\ s^{-1}\ sr^{-1}\ keV^{-1}}$.
\item The column density accounting for the MW and CXB absorption is based on the dust emission map from the \textit{Planck} survey, rather than the \ion{H}{1} map.
\item We include the Fe-L line as another Gaussian function in our model. Additionally, we disable strong iron emission lines in the 0.7--1.0 keV band in the APEC model.
\item We determine our measurements and uncertainties using the MCMC method, as opposed to the $\chi^2$-statistic method used by \citetalias{henley2012xmm}.
\end{enumerate}

\begin{figure}[t]
\centering
\includegraphics[width=0.472\textwidth]{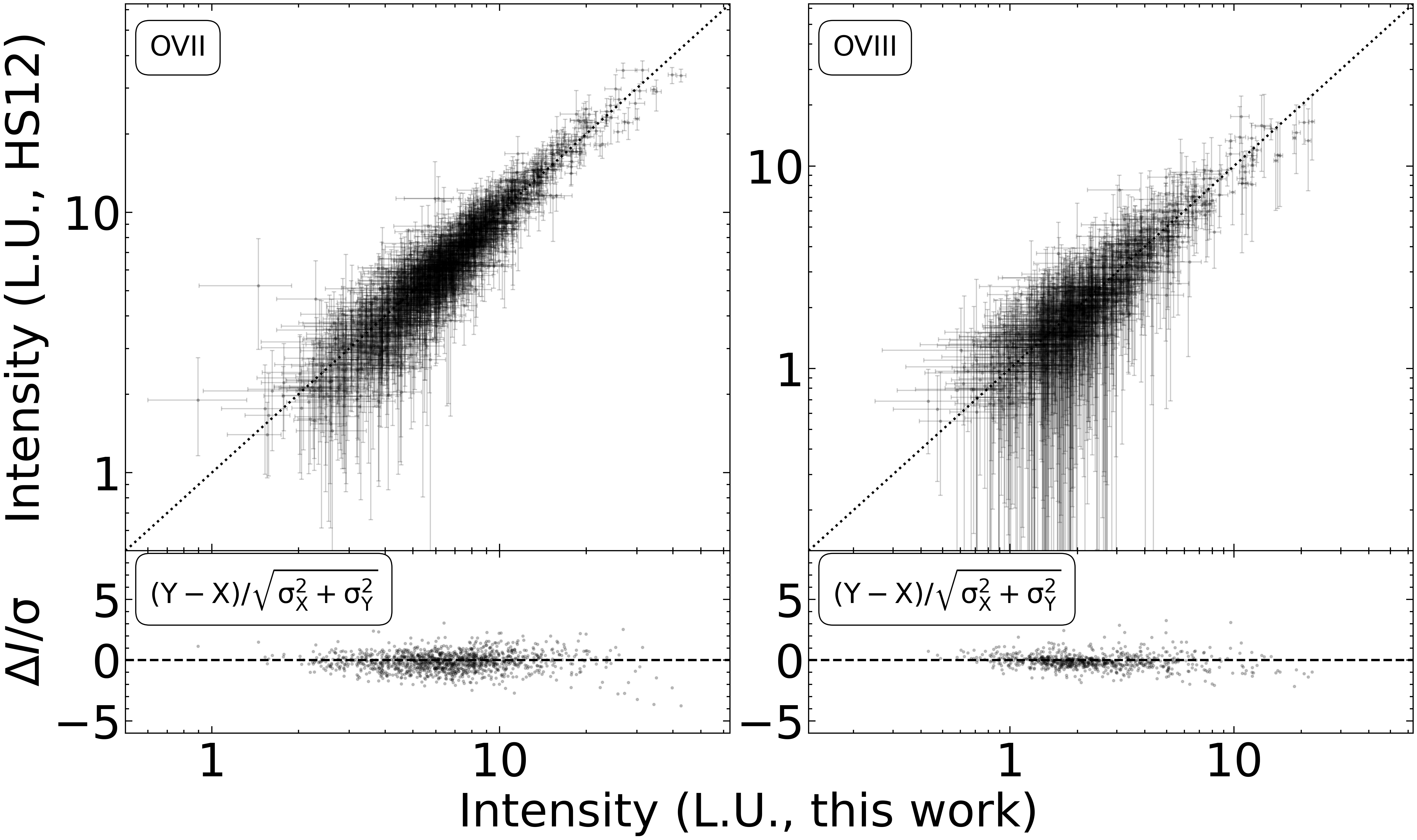}
\caption{Top row: comparison of shared \ion{O}{7} and \ion{O}{8} intensity measurements between this work and \citetalias{henley2012xmm}. Bottom row: weighted differences derived from intensity difference ($\Delta I$) divided by total uncertainty ($\sigma$). Interpretation of the differences is in Section \ref{Data Comparison}.}
\label{fig:comparison}
\end{figure} 

To visualize the impact of these differences, we conduct a comparative analysis using 1312 shared observations between our study and \citetalias{henley2012xmm}.
In Figure \ref{fig:comparison}, we show all shared measurements with ours on the X-axis and \citetalias{henley2012xmm}'s on the Y-axis. 
The figure shows that our measurements largely align with those of \citetalias{henley2012xmm}, with \ion{O}{7} and \ion{O}{8} medians 0.19 and 0.15 L.U. higher than \citetalias{henley2012xmm}'s.

The discrepancy in \ion{O}{8} is most likely caused by the metal abundance table adopted in the spectral fitting.
In our model, the MW CGM is modeled using an APEC model with \ion{O}{8} lines ($\approx$ 654 eV) disabled, but the \ion{O}{7} K$\beta$ line ($\approx$ 666 eV) remains.
Shifting from the \cite{anders1989abundances} to the \cite{lodders2003solar} abundance table nearly halves the oxygen abundance (i.e, from $8.51\times10^{-4}$ to $4.90\times10^{-4}$). This change results in a weaker \ion{O}{7} K$\beta$ intensity in the MW APEC model, which can lead to a slight increase in the measured intensity of the \ion{O}{8} line.
This is tested by randomly fitting 500 SXRB spectra using both abundance tables while keeping other settings unchanged.
The results show that $\approx$ 80\% of the \ion{O}{8} values are higher when the \cite{lodders2003solar} table is used, supporting this explanation.

In conclusion, our \ion{O}{7} and \ion{O}{8} measurements generally align with those of \citetalias{henley2012xmm} with minor differences due to methodological differences.

\section{Temporal Variation in SWCX}
\label{sec:SWCX_variation}

\blue{Studying the temporal variation of SWCX helps us understand its characteristics and accurately derive the intrinsic MW emission by subtracting this variation.}

Hot gas emissions from the LHB and MW are expected to remain constant over decades. Therefore, any temporal variations in a given direction are attributed to SWCX \citep[e.g.,][]{kuntz2019solar}.

The SWCX emission in the SXRB has two primary contributors, the magnetospheric SWCX and the heliospheric SWCX.
The former arises from interactions between highly ionized species in the solar wind with the neutral gas in Earth's atmosphere and typically exhibits variations on timescales of hours to days (see Section \ref{MagSWCX} for a detailed discussion).
The heliospheric SWCX occurs when the solar wind interacts with the neutral ISM entering the heliosphere and generally varies over the solar cycles \citep{cravens2001temporal}. Notably, in certain directions such as those parallel to the Parker spiral, the variation timescales can be shorter than a week \citep{kuntz2019solar}. \cite{qu2022solar} reported a temporal variation of the observed \ion{O}{7} and \ion{O}{8} lines over a solar cycle using \xmm data from \citetalias{henley2012xmm}. This variation is highly correlated with the solar cycle and is likely associated with the heliospheric SWCX. 
The 22-year data presented in this study allows a more precise investigation of this long-term variation over an entire solar magnetic cycle, which will be further investigated in future work.

\subsection{Characterizing the Long-Term SWCX}
\label{sub:Helio SWCX Method}

\begin{figure}[t!]
\centering
\includegraphics[width=0.475\textwidth]{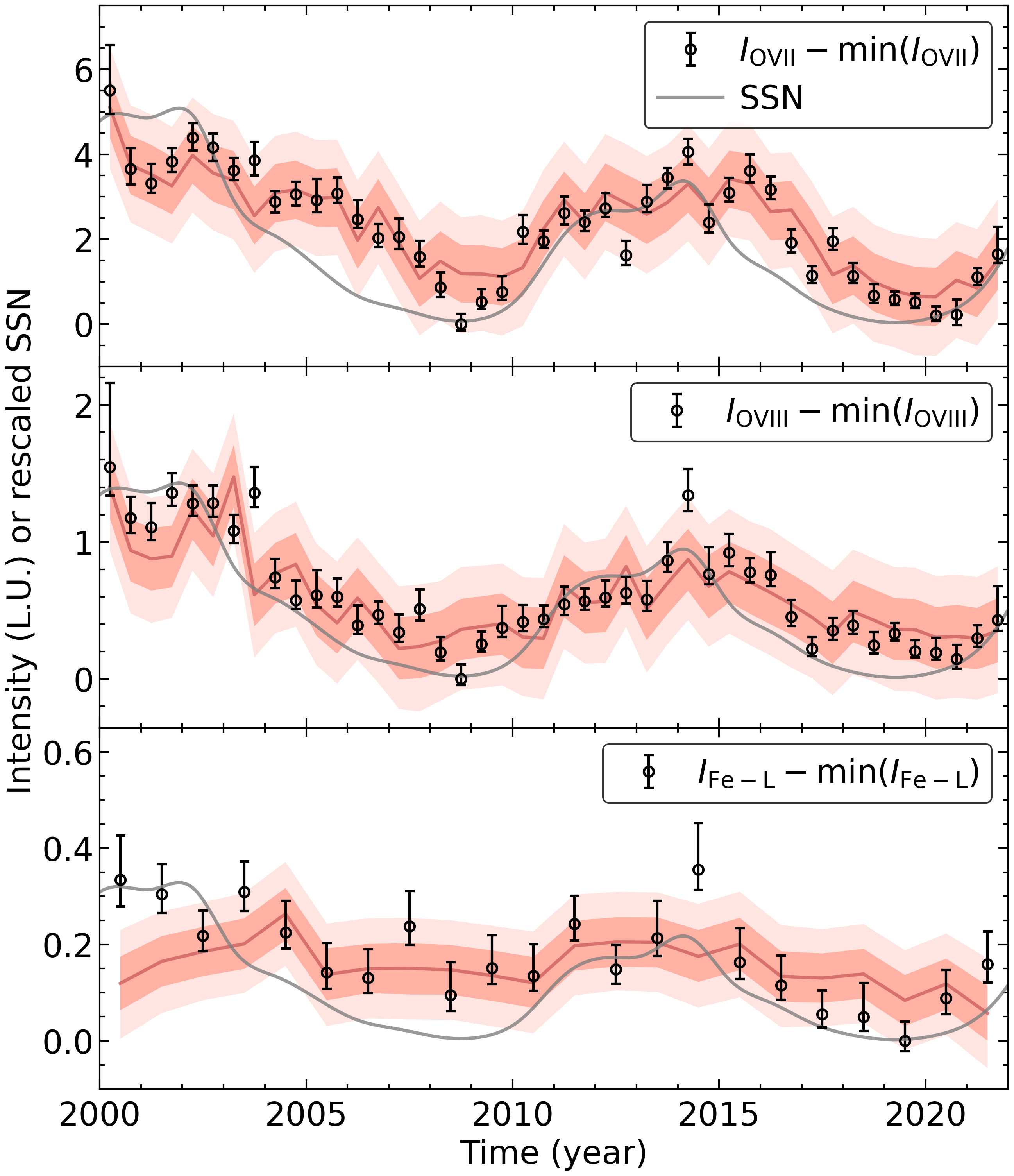}
\caption{Long-term variations in the observed \ion{O}{7}, \ion{O}{8}, and Fe-L line intensities from 2000 to 2022. Sunspot numbers (SSN) are smoothed and \red{rescaled individually} for better visualization. The circles represent intensities in half- or 1-year bins using the clean sample, \red{adjusted by subtracting minimum values of 3.20, 0.54, and 0.46 L.U., respectively.} The reddish curves represent the medians, $1\sigma$, and $2\sigma$ uncertainties derived from the ``close-pair'' method. The variations in \ion{O}{7} and \ion{O}{8} intensities correlate with the solar cycle, suggesting a strong association with SWCX. \red{Moderate Fe-L correlation indicates that Fe-L is not primarily produced by SWCX (see text for details).}}
\label{fig:four_var}
\end{figure}

\red{The \textit{XMM-Newton's} uneven pointing can introduce observational biases. These biases are mitigated in half-year or longer intervals, as \textit{XMM-Newton} observes the brighter areas of the sky twice a year. This scheduling enables us to investigate long-term variations by plotting the median intensities over half- or 1-year intervals using the clean sample, as illustrated by the dots in Figure \ref{fig:four_var}. Although the results reveal clear trends correlating with solar activity, a more robust approach independent of pointing biases is necessary to confirm the authenticity of the observed variations.}

\begin{figure*}[t!]
\centering
\includegraphics[width=0.82\paperwidth]{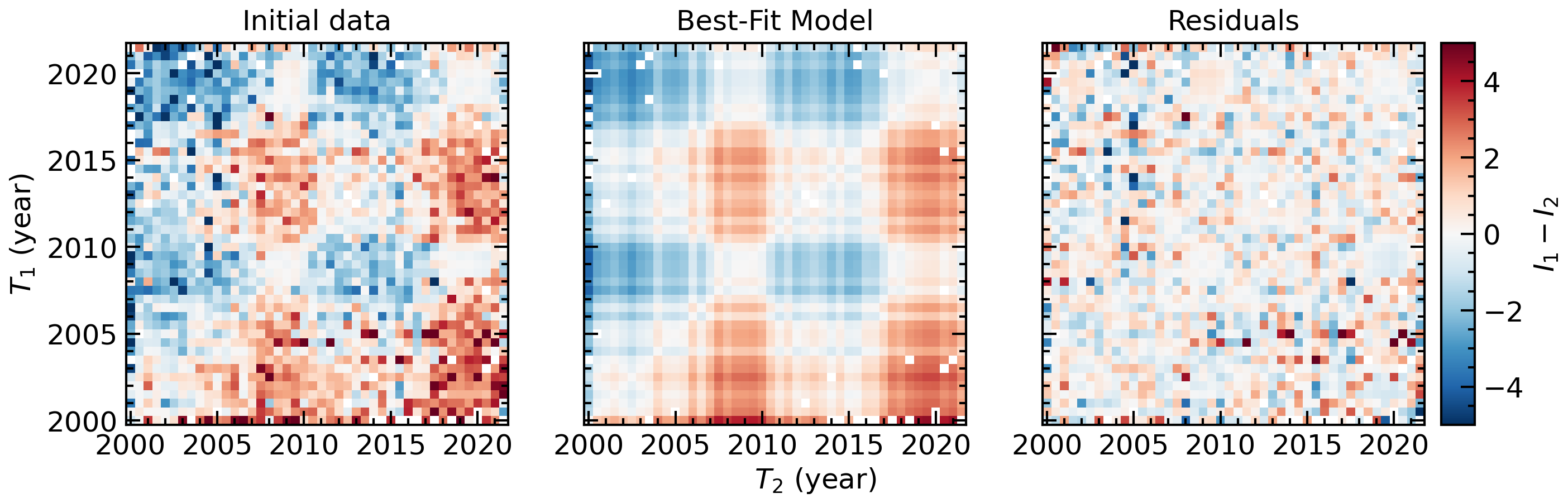}
\caption{Example of MCMC analysis to derive long-term variations in \ion{O}{7} intensity from close pairs. Left: the \blue{medians} of \ion{O}{7} intensity differences in half-year intervals. Middle: the intensity differences reconstructed after MCMC fitting. Right: residuals. The MCMC fitting process and results are detailed in Section \ref{sub:Helio SWCX Method} and Section \ref{sub:Helio SWCX result}.}
\label{fig:O7rebin}
\end{figure*}

Here, we apply a ``close-pair'' method that utilizes observation pairs projected within $\leq 2^\circ$, since the MW emission shows a strong correlation within 2$\degr$ (e.g., \citealt{kaaret2020disk}; \citetalias{Qu:2023aa}).
\blue{By only considering the clean sample}, we identify over 26000 close pairs and calculate their intensity differences (i.e., $I_{\rm 1, O VII}-I_{\rm 2, O VII}$). They are then grouped into half-year intervals based on the observation dates. The median values for each interval are shown in the left panel of Figure \ref{fig:O7rebin}. For better visualization, each pair is counted twice by inverting the order.

We reconstruct the temporal variation of the SWCX using the difference calculated based on close pairs.
In this fitting, we only use pairs with the observation date $T_1$ later than $T_2$, meaning only pairs above the diagonal in Figure \ref{fig:O7rebin}.
We then adopt a non-parametric model to fit these re-binned differences in the Bayesian framework using the \texttt{emcee} implementation (\citealt{Foreman-Mackey:2013aa}). It has a likelihood function of
\begin{equation}
\begin{aligned}
\ln p &= -\frac{1}{2} \sum \left[ \frac{(y - y_{\rm m})^2}{\sigma_{\rm m}^2 + \sigma_{\rm p}^2} + \ln(\sigma_{\rm m}^2 + \sigma_{\rm p}^2) \right], 
\end{aligned}
\end{equation}
where $y$ and $\sigma_{\rm m}$ are the observed difference and uncertainty.
\red{For O VII and O VIII, the model $y_{\rm m}$ represents the difference calculated based on a step function with 44 free parameters, characterizing the SWCX variations in each half-year bin from 2000 to 2022. In contrast, the long-term variation in Fe-L intensity is modeled with 22 parameters corresponding to 1-year intervals, due to its lower signal-to-noise ratio.} The parameter ${\sigma }_{\rm p}$ is an empirical parameter to account for additional scatter in data, which is different for \ion{O}{7}, \ion{O}{8}, and Fe-L. Specifically, the fitted values are $0.67 \pm 0.03$, $0.22 \pm 0.01$, and \red{$0.03 \pm 0.01$}, respectively.
The MCMC sampling assigns 150 walkers, each taking 30,000 steps. Only the last 10\% of the chain is used to derive the results for best-fit parameters. 
Figure \ref{fig:O7rebin} shows the best-fit model and residuals of the \ion{O}{7} difference, for example.

\subsection{The Empirical SWCX Model}
\label{sub:Helio SWCX result}

The results derived from the \blue{``close-pair'' method, along with the regrouped median intensities are shown in Figure \ref{fig:four_var}. The fitted model is represented by reddish curves with medians, $1\sigma$, and $2\sigma$ intervals outlined.} 
The gray curves represent the re-scaled and smoothed sunspot number (SSN), as a direct tracer for solar activity. 
The figure shows that \ion{O}{7} and \ion{O}{8} emissions rise with an increase in SSN and decrease when solar activity diminishes.
\red{In contrast, this trend is less prominent for Fe-L emissions.}

To evaluate the consistency between the ``close-pair'' and ``rebin'' approaches, we calculate the normalized differences, $(I_{\rm cp}-I_{\rm rebin})/\sigma$. \red{The result shows that $\sim 80\%$ of the \ion{O}{7} differences and $\sim 70\%$ of both \ion{O}{8} and Fe-L differences fall within $\pm 1$, indicating a high consistency between the two approaches.}

We examine the relations between the SSN and each line emission by applying the Pearson correlation coefficient test on the median intensities of the three lines and the median SSNs. The calculated correlation coefficients are 0.85, 0.83, and \red{0.44}, respectively, with corresponding $p$-values of $2.43 \times 10^{-13}$, $2.17 \times 10^{-12}$, \red{$4.00 \times 10^{-2}$}. These $p$-values represent the probabilities that the observed relationships occurred by chance. The results suggest strong positive correlations between SSN and the \ion{O}{7} or \ion{O}{8} intensities, while the correlation with Fe-L emission is \red{moderate}.

\red{In addition, Figure \ref{fig:four_var} shows long-term variations in \ion{O}{7}, \ion{O}{8}, and Fe-L intensities with medians of 1.8, 0.3, and 0.1 L.U., against observed medians of 6.2, 1.4, and 0.8 L.U., respectively. This implies SWCX contributes roughly 30\%, 20\%, and 10\% on average to the observed intensities, significantly impacting \ion{O}{7} and \ion{O}{8} emissions, with a more modest effect on Fe-L emission.
Moreover, the \ion{O}{7} and \ion{O}{8} variations can reach up to approximately 4.5 L.U. and 1.3 L.U., consistent with \cite{henley2010xmm}.}

\section{Soft X-ray Emission of the MW}
\label{sec:Soft X-ray distribution of the MW}

The MW's intrinsic emission is contaminated by foreground SWCX and LHB emissions, and partially absorbed by the medium along the line of sight. In the following, we quantify and correct these factors to derive the true MW emission.

The LHB is known to contribute more to the 1/4 keV band than the 3/4 keV band \citep{Kuntz:2000}, implying it produces more \ion{O}{7} emission than \ion{O}{8} emission. Shadowing observations towards the foreground molecular cloud, MBM 12, reveal an observed \ion{O}{7} emission of $1.8^{+0.5}_{-0.6}$ L.U., with about 0.28 L.U. expected from the LHB \citep{koutroumpa:2011}. Further modeling of the observed \ion{O}{7} emission, accounting for multiple components including the LHB, suggests that the LHB's contribution is $<1$ L.U. \citep[e.g.,][]{miller2015constraining, wulf2019high}. In addition, we estimate the LHB emission based on the emission measure, which is typically less than $7\times10^{-3}$ $\rm cm^{-6}$ pc \citep{liu2016structure}, and $\sim 2 \times 10^{-3}$ $\rm cm^{-6}$ pc at medium Galactic latitudes ($|b| \sim 20 \degr$; \citealt{yeung2023srg}). Assuming a temperature of 0.084 keV, the expected \ion{O}{7} and \ion{O}{8} emissions from the LHB are under 0.6 and $8.4 \times 10^{-4}$ L.U., respectively. 
These findings indicate that the \ion{O}{7} emission from the LHB is generally less than 1 L.U., contributing a maximum of 17\% to the median observed \ion{O}{7} intensity ($\approx 6 \ \rm L.U.$). Since the LHB's contribution is relatively small and varies with direction, we have deferred such detailed analysis to future research.

To remove the SWCX contribution, we subtract the long-term variation of each line shown in Figure \ref{fig:four_var} from the observed intensities. \red{Although Fe-L shows only a moderate correlation with solar activity, we include it in the correction for consistency.}

\begin{figure}[t!]

    \centering
    
    \begin{minipage}[ht!]{0.475\textwidth}
        \includegraphics[width=\linewidth]{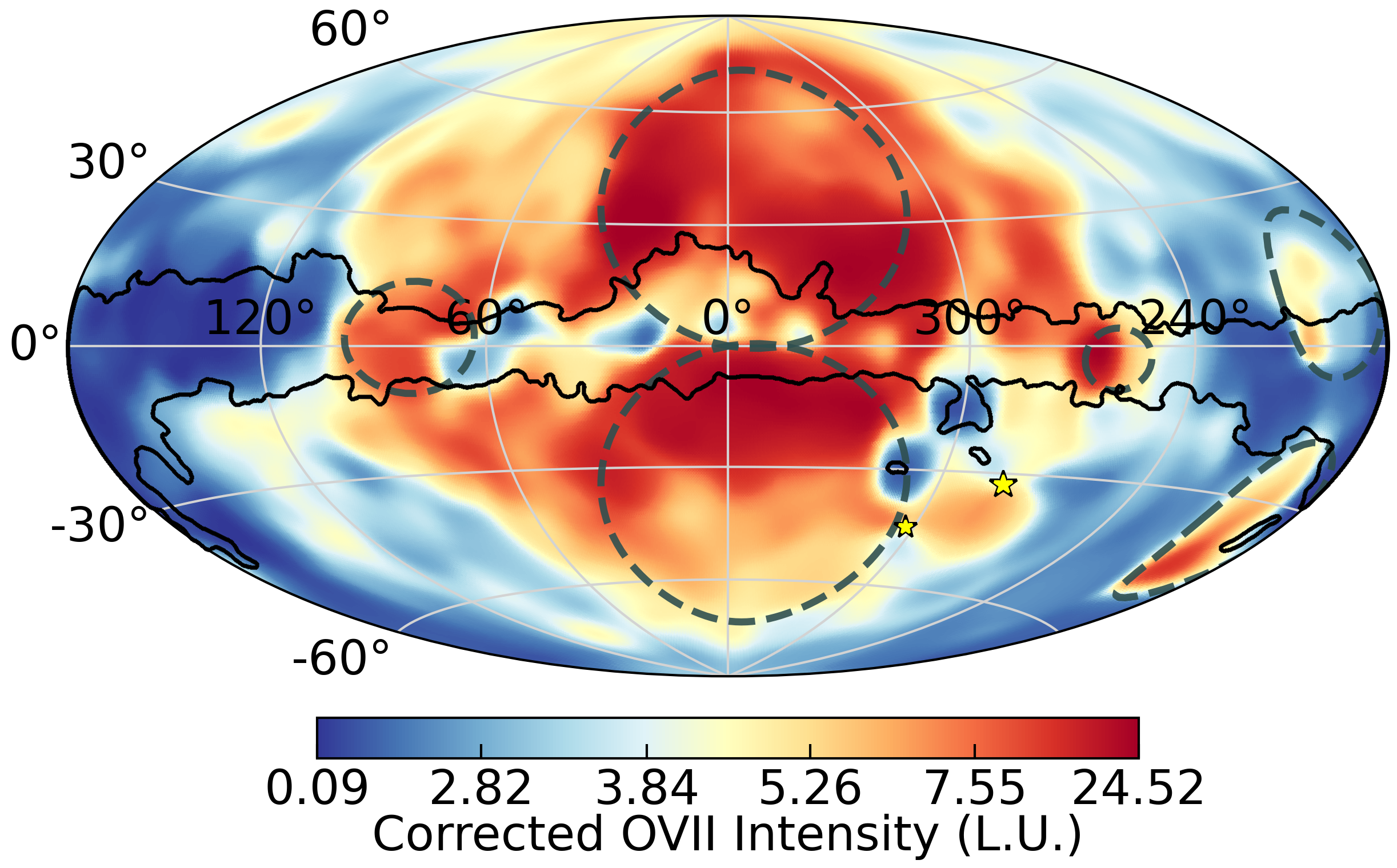}
    \end{minipage}
    \hfill
    \begin{minipage}[ht!]{0.475\textwidth}
        \includegraphics[width=\linewidth]{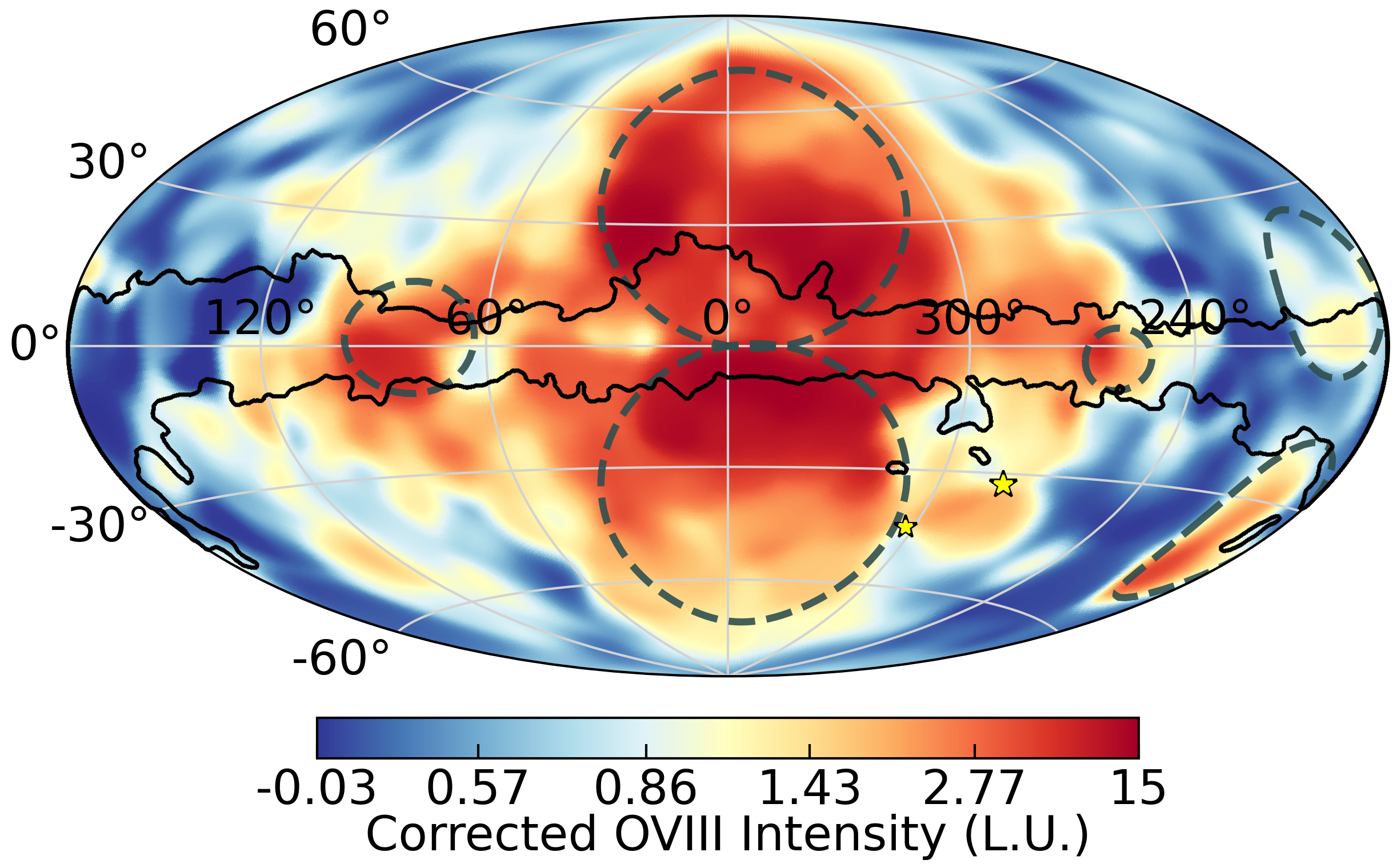}
    \end{minipage}
    
    \begin{minipage}[ht!]{0.475\textwidth}
        \includegraphics[width=\linewidth]{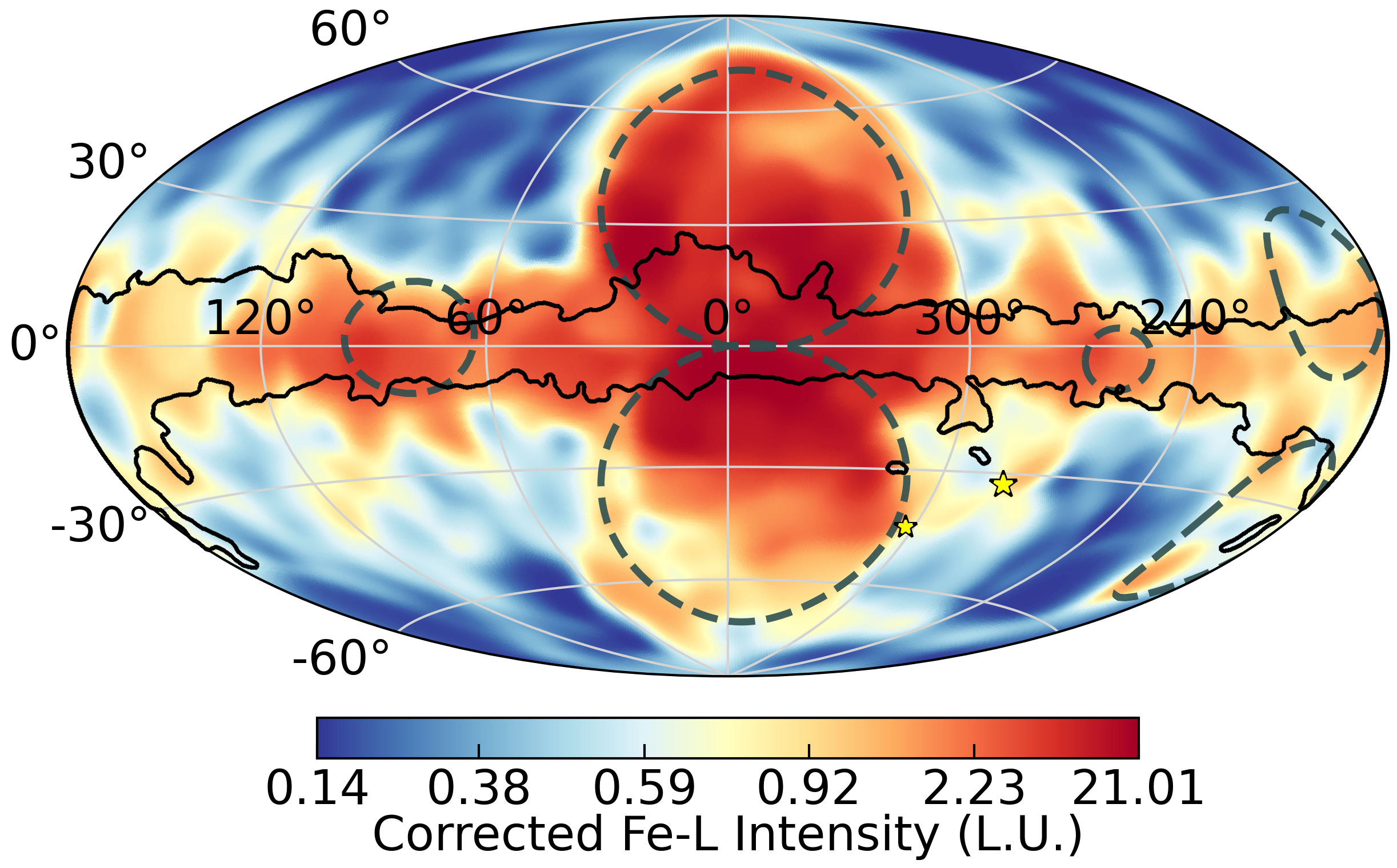}
    \end{minipage}
\caption{All-sky maps depicting the MW soft X-ray emissions. From top to bottom: SWCX- and absorption-corrected \ion{O}{7}, \ion{O}{8}, and Fe-L maps.
These maps are centered on the Galactic center and smoothed with a Gaussian function of $\sigma =4\degr$. \red{Dashed lines indicate large-scale structures introduced in Figure \ref{fig:observedAllsky}. Black outlines hydrogen-dense areas where $N_{\rm H, dust} > 5 \times 10^{21}$ $\rm cm^{-2}$.} The oxygen line emissions extend beyond the eROSITA bubbles (eRBs) and decrease from the Galactic center. The Fe-L line emission closely matches the eRBs and Galactic disk.
The emission intensity is proportion to $n_e^{2} \epsilon (T) Z$, meaning a higher intensity suggests an increased plasma density, metallicity, or emissivity at temperatures optimal for emission (i.e., $\approx 2\times10^6$ K, $\approx 3\times10^6$ K, and $\approx7\times10^6$ K for \ion{O}{7}, \ion{O}{8}, and Fe-L). The LHB emission is not corrected and may contribute to these line emissions, particularly for \ion{O}{7} (see Section \ref{sec:Soft X-ray distribution of the MW}).}
\label{fig:ALLSKYmaps}
\end{figure}

After subtracting the foreground from the observed emission, the unabsorbed MW emission can be calculated as
\begin{equation}
\begin{aligned}
I_{\rm MW} = (I_{\rm obs} - I_{\rm fg}) \, e^{\tau_{\rm eff}},
\end{aligned}
\end{equation}
\red{where $e^{\tau_{\rm eff}}$ serves as the absorption correction factor with $\tau_{\rm eff}$ being the effective optical depth. $\tau_{\rm eff}$ is determined by $\sigma N_{\rm H, eff}$, where $\sigma$ is the absorption cross-section \citep[][]{balucinska1992photoelectric, yan1998photoionization} and $N_{\rm H, eff}$ is the effective column density for a given line of sight.}

In the spectral modeling, $N_{\rm H, eff}$ is simply the total $N_{\rm H}$ along each observation's line of sight, derived from the dust emission map ($N_{\rm H, dust}$, Section \ref{subsec:gas absorption}). This approach assumes that both the CGM and CXB emissions are located behind all absorbing layers. This assumption is commonly used \citep[e.g.,][]{henley2012xmm, miller2015constraining} because most MW hot CGM emissions are situated beyond the MW disk, a region where the majority of absorbing material lies. However, this approach overestimates the absorption correction in high $N_{\rm H, dust}$ regions, particularly in the MW disk. Because the X-ray-emitting gas is mixed with the absorbing medium \red{rather than being entirely behind it}.

\red{Determining the effective absorption factor for these regions requires understanding the spatial distributions of both the high-column-density cool gas and X-ray-emitting hot gas. However, a more sophisticated model that considers both absorption and emission simultaneously is beyond the scope of this paper.}

\red{Here, we empirically derive the maximum-allowed absorption factor mainly for high $N_{\rm H, dust}$ observations, based on data quality.}
\red{First, we assume that the absorbing medium and the X-ray-emitting gas are cospatial and the source function ($S$) remains constant across all absorption optical depths. 
In this scenario, the amount that escapes from the absorbing region is proportional to the integral $\int_0^{\tau} e^{-\tau}~ {\rm d}\tau$, or equivalently, $1-e^{-\tau}$.
We then derive a maximum optical depth ($\tau_{\rm max}$) by assuming $e^{-\tau_{\rm max}} = \sigma_I/I$, where $I$ and $\sigma_I$ are the measured line intensity and corresponding uncertainty for each observation. The calculated $\tau_{\rm max}$ represents the maximum optical depth allowed by the data quality as emissions beyond this threshold are obscured by the measurement uncertainty. Therefore, absorption correction is only applied to nearby gas with optical depths within $\tau_{\rm max}$.
Considering the cospatial absorbing and emitting gas, the unabsorbed and observed intensities can be calculated as $S\tau_{\rm max}$  and $S(1-e^{\tau_{\rm max}})$. Thus, the maximum-allowed absorption factor is derived as $\tau_{\rm max}/(1-e^{-\tau_{\rm max}})$.
} 

For each observation, we always use the lower value between the dust-inferred and the maximum-allowed absorption factor. Consequently, $\approx 10\%$ of the observations are corrected using the later factor, primarily concentrated in the Galactic disk region.

We note that using the data-limited factor primarily accounts for nearby gas absorption, which might lead to an underestimation of total absorption. Additionally, selecting the lower of two distinct correction factors may introduce inconsistencies, especially where these methods significantly diverge in different galactic regions. However, considering our data limitations, this is likely the most suitable approach we can adopt.

Upon applying SWCX and absorption corrections, we applied Gaussian smoothing to the data. Recent studies hint at the presence of small-scale temperature structures in the MW's hot gas, revealing significant variations at approximately $4\degr-6\degr$ angular scales 
(\citealt{kaaret2020disk}; \citetalias{Qu:2023aa}). As such, we choose to apply the smoothing with a standard deviation of $4\degr$ to enhance the visibility of these variations in all-sky maps.
\red{The smoothed all-sky maps are displayed in Figure \ref{fig:ALLSKYmaps}, with large-scale structures (dash lines) highlighted. Hydrogen-dense areas with $N_{\rm H, dust} > 5 \times 10^{21}$ $\rm cm^{-2}$ are outlined in black. Emissions from these areas are typically corrected using the maximum-allowed absorption factor.} This plot shows that the \ion{O}{7} and \ion{O}{8} intensities decrease away from the Galactic center, tracing the hot CGM at $\approx 2\times 10^6$ K and $\approx 3\times 10^6$ K.

\red{In contrast, the Fe-L emission matches closely with the eROSITA bubbles (eRBs) and Galactic disk, implying two different origins of this hotter plasma ($\approx 7\times 10^6$ K).}
Previous studies have noticed such plasma in eRBs \citep[e.g.,][]{das2019multiple, ponti2022abundance} and can be explained by significant energy injections from the Galactic center \citep{Kataoka:2018}.
However, the origin of the Fe-L disk remains controversial. 
A possible contributor to this emission is M dwarfs, the most abundant stars ($\sim75\%$) in the solar neighborhood \citep{Reyle:2021}.
Despite their low mass ($\approx 0.1-0.6 \ M_{\odot}$) and luminosity, M dwarfs generally exhibit higher soft X-ray luminosity compared to the Sun due to more flare activities \citep[e.g.,][]{Osten:2005, Wargelin:2008}. 
Observations targeting $l,b=165\degr,5\degr$ indicate excess emissions in the Fe-L band, which can be explained by including an extra spectral emission model of M dwarfs \citep{wulf2019high}.
A more recent study of the eFEDS field ($l \sim 220\degr-230\degr$ and $b \sim 20\degr-40\degr$) also observes such excess emissions, which can be well-fitted using an additional APEC model with $kT\sim 0.7$ keV \citep{ponti2022abundance}. Including the same stellar model from \cite{wulf2019high}, they find that the stellar contribution accounts for a significant portion, but not all. They propose that the rest of the excess emissions are from the Galactic corona, a plasma phenomenon likely arising from hot outflows or Galactic fountains originating in the Galactic disk.

As the contributions from different sources remain under debate, our Fe-L emission data could be crucial for future explorations in this area.

\section{Discussion}
\label{sec:Discussion}

Using the X-LEAP sample, we characterize the long-term variation in \ion{O}{7} and \ion{O}{8} emissions, and construct SWCX- and absorption-corrected all-sky maps for all three emission lines of interest. In this section, we explore further implications of the X-LEAP sample, including using the $E_{\rm O VII}$ as a solar activity tracer, evaluating the impact of magnetospheric SWCX on short-term variations, investigating the potential spatial dependence of SWCX on ecliptic latitude, and discussing future work.

\subsection{\ion{O}{7} Line Centroid as an SWCX Tracer}
\label{subsec:O7 centroid}

\begin{figure}[t!]
\centering
\includegraphics[width=0.39\paperwidth]{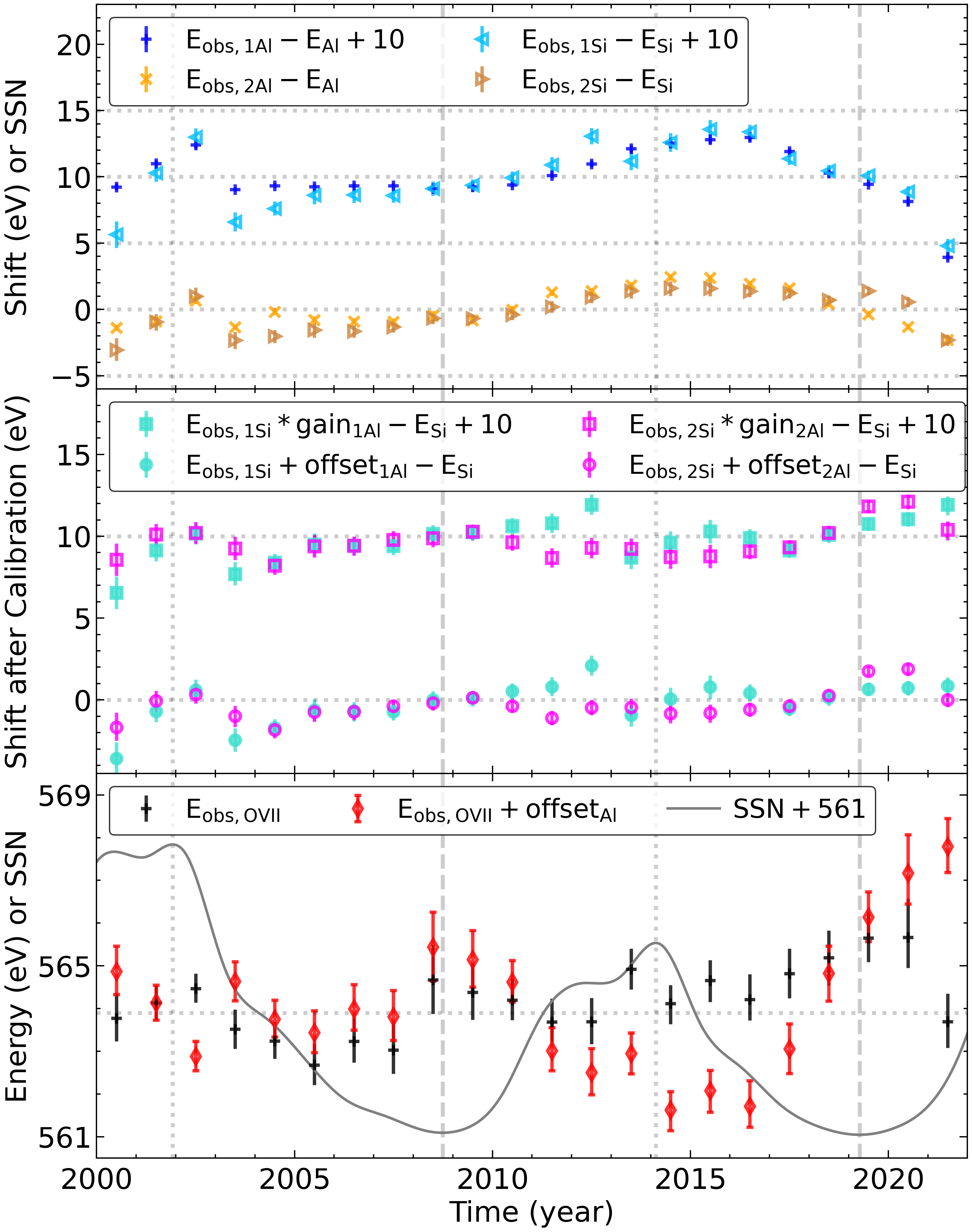}
\caption{Top: yearly shifts in observed Al and Si instrumental line centroids in both MOS1 and MOS2. These shifts are with an uncertainty of 5 eV, likely due to temperature variations in the instrument. Middle: calibrated Si line centroids derived from Al-inferred gain or offset adjustments. A smaller total $\chi^2$ in the offset case suggests a closer relation between the shift and offset.
Bottom: the observed and corrected $E_{\rm OVII}$ over time. The correction is done using the Al-inferred offset results to minimize the instrument-induced variations. The resulting trend suggests a potential anti-correlation with solar activity, implying $E_{\rm OVII}$ as a viable tracer of solar activity, even with limited spectral resolution.}
\label{fig:EO7}
\end{figure}

The \ion{O}{7} line is a triplet consisting of a resonance line, an inter-combination line, and a forbidden line, as detailed in Section \ref{threelines}.
The resonance line dominates the MW hot gas emission due to efficient collisional excitation, while the forbidden line dominates in SWCX emission because of the downward electron cascade favoring triplet states (see \citealt{porquet2010he} for a detailed review).  
Thus, the $E_{\rm OVII}$ can theoretically be used to trace the strength of SWCX.

The EPIC-MOS detectors have a spectral resolution of $\approx40$ eV at 500 eV, unable to resolve the resonance and forbidden lines separated by $\approx13$ eV.
However, shifts in the $E_{\rm OVII}$ may still be detectable.
\cite{ponti2022abundance} detects an $E_{\rm OVII}$ shift towards the forbidden line when the SWCX component dominates using SRG/eROSITA, which has a spectral resolution comparable to \xmm's \citep{predehl2021erosita}.

To examine $E_{\rm OVII}$ shifts in MOS1 and MOS2, we re-sample the observed \ion{O}{7} line centroids, together with the instrumental line centroids in 1-year bins using the clean sample.
Interestingly, both instrumental centroids shift similarly from their rest energies over time (top panel, Figure \ref{fig:EO7}), likely due to temperature variations in the MOS detectors \citep{Abbey:2003}. According to the \xmm technical notes\footnote{https://xmmweb.esac.esa.int/docs/documents/CAL-TN-0018.pdf}, line centroid shifts are corrected by assuming a linear relationship between the rest and observed energies:
\begin{equation}
\begin{aligned}
E_{\rm rest} = {\rm gain} \times E_{\rm obs} + {\rm offset}.
\end{aligned}
\end{equation}

However, after correction, line energy can still vary with an uncertainty of 5 eV over the full energy range for EPIC-MOS$^9$, consistent with the observed shifts.
Due to limited spectral resolution, remaining gain and/or offset changes can not be accurately determined. To evaluate the more dominant factor, we first assume that these shifts are solely due to either gain or offset changes. We then use the Al-inferred gain and offset (i.e., $\rm gain$ = $E_{\rm{Al}}/E_{\rm{obs, Al}}$ and $\rm{offset}$ $=E_{\rm{Al}} - E_{\rm{obs, Al}}$) to calibrate the observed Si line centroids (middle panel, Figure \ref{fig:EO7}). \red{The results suggest a smaller total $\chi^2$ and lower dispersion (i.e., $\sigma = 1.04$ eV compared to 1.09 eV) in the offset case}, implying a tighter association between the shift and offset. Thus, we correct the $E_{\rm obs, OVII}$ using the Al-inferred offset results to minimize the instrument-induced variations.   
The bottom panel of Figure \ref{fig:EO7} presents both the observed and corrected $E_{\rm OVII}$ over time. The results show a potential anti-correlation with solar activity, especially during solar maxima and minima.

\blue{The findings suggest that the $E_{\rm OVII}$ effectively traces solar activity, even with limited spectral resolution.
Future X-ray missions like HUBS and LEM \citep{HUBS, LEM} will have higher spectral resolutions capable of resolving the \ion{O}{7} triplet lines.}
This improvement will allow the use of line ratios between the forbidden and resonant lines to assess solar activity levels, or even distinguish the SWCX and MW emissions directly. 

\subsection{Evidence of Magnetospheric SWCX}
\label{MagSWCX}
\begin{figure}[t!]
\centering
\includegraphics[width=0.40\paperwidth]{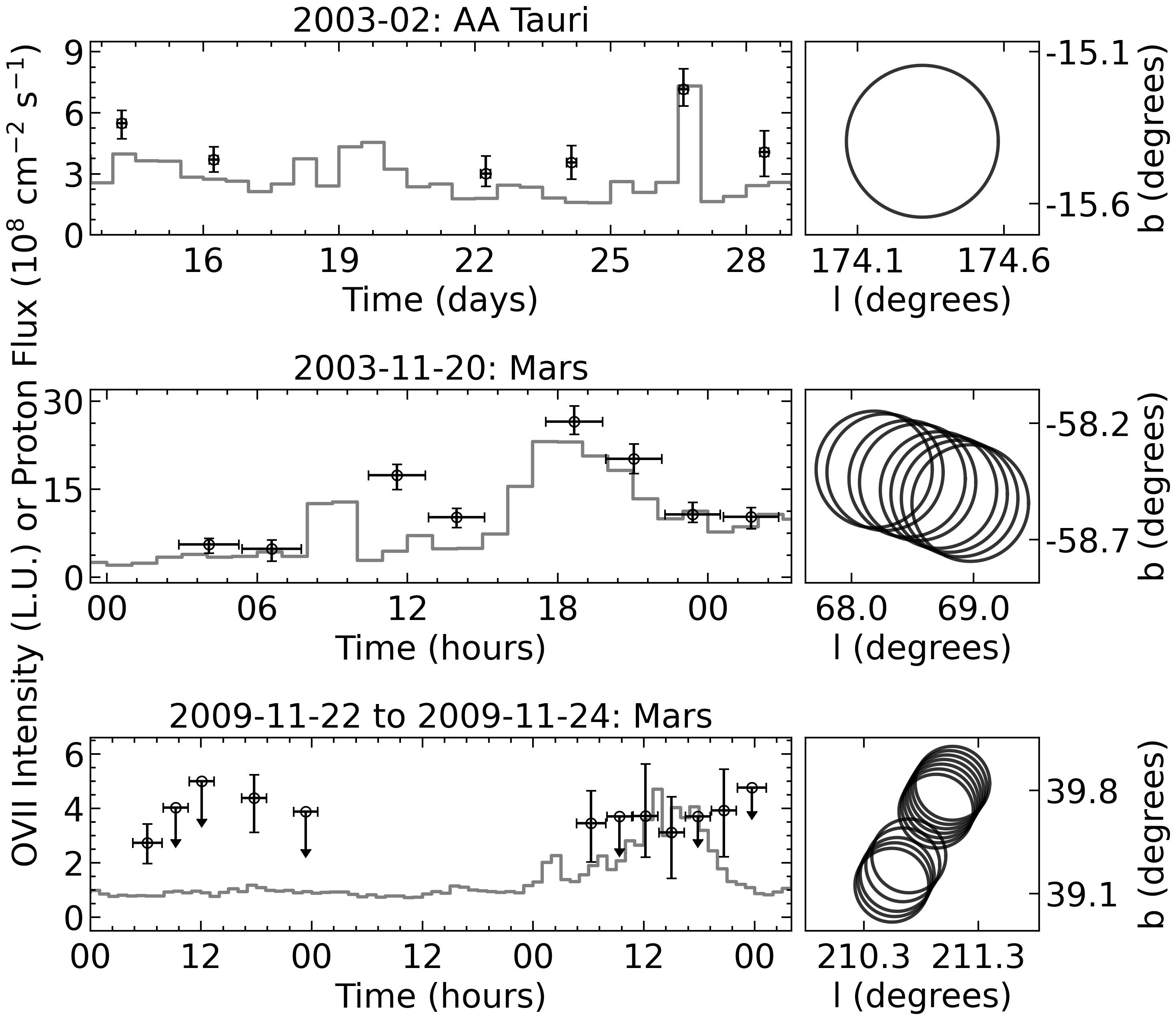}
\caption{Left column: \ion{O}{7} intensity measurements/upper limits obtained from observations of the same targets within one month, one day, and two days, respectively. The gray lines represent the 12-hour and 1-hour averaged proton fluxes retrieved from ACE. Right column: The corresponding FOVs for these observation sets. The AA Tauri observations maintain the same direction, while the Mars observations shift by $\approx 0.5\degr$. Significant variations in the top two rows are likely associated with the magnetospheric SWCX.}
\label{fig:samedir}
\end{figure}

The magnetospheric SWCX is sensitive to changes in solar wind conditions and Earth's magnetic field \citep{ishikawa2013suzaku}, \blue{which may lead to observable intensity variations over short periods.
To investigate this, we identify same-direction observations in our dataset and find two representative observation sets of Mars \citep{dennerl2006first} and AA Tauri, a young variable star at 150 pc \citep{loomis2017multi}.}
Both objects are effectively masked out by \texttt{cheese} as point sources, and their FOVs remain almost unchanged.
Therefore, any observed intensity changes are independent of both the target emissions and changes in observation direction.

Figure \ref{fig:samedir} shows the observed variations in \ion{O}{7} intensities of the Mars and AA Tauri observations, along with time-averaged solar proton flux data (grey line) retrieved from the Advanced Composition Explorer\footnote{https://izw1.caltech.edu/ACE/ASC/level2/index.html} (ACE). 
In the top two panels, the \ion{O}{7} intensities exhibit significant intensity enhancements of $\approx$ 5 L.U. within a month and $\approx$ 23 L.U. within a day. These magnitudes of increase are within those reported by others \citep[e.g.,][]{snowden2004xmm, fujimoto2007evidence, koutroumpa2007ovii}. Additionally, intensity peaks correlate with periods of maximum proton flux, suggesting that they are induced by rapid changes in solar activity.
In contrast, the 2009 Mars observations (bottom row) display stable intensities, unsurprising given the proton flux is lower than in 2003.

These findings indicate that short-term variations in \ion{O}{7} intensity are primarily driven by changes in solar activity. 
This highlights the need to account for magnetospheric SWCX in future X-ray studies.

\begin{figure}[t!]
\centering
\includegraphics[width=0.39\paperwidth]{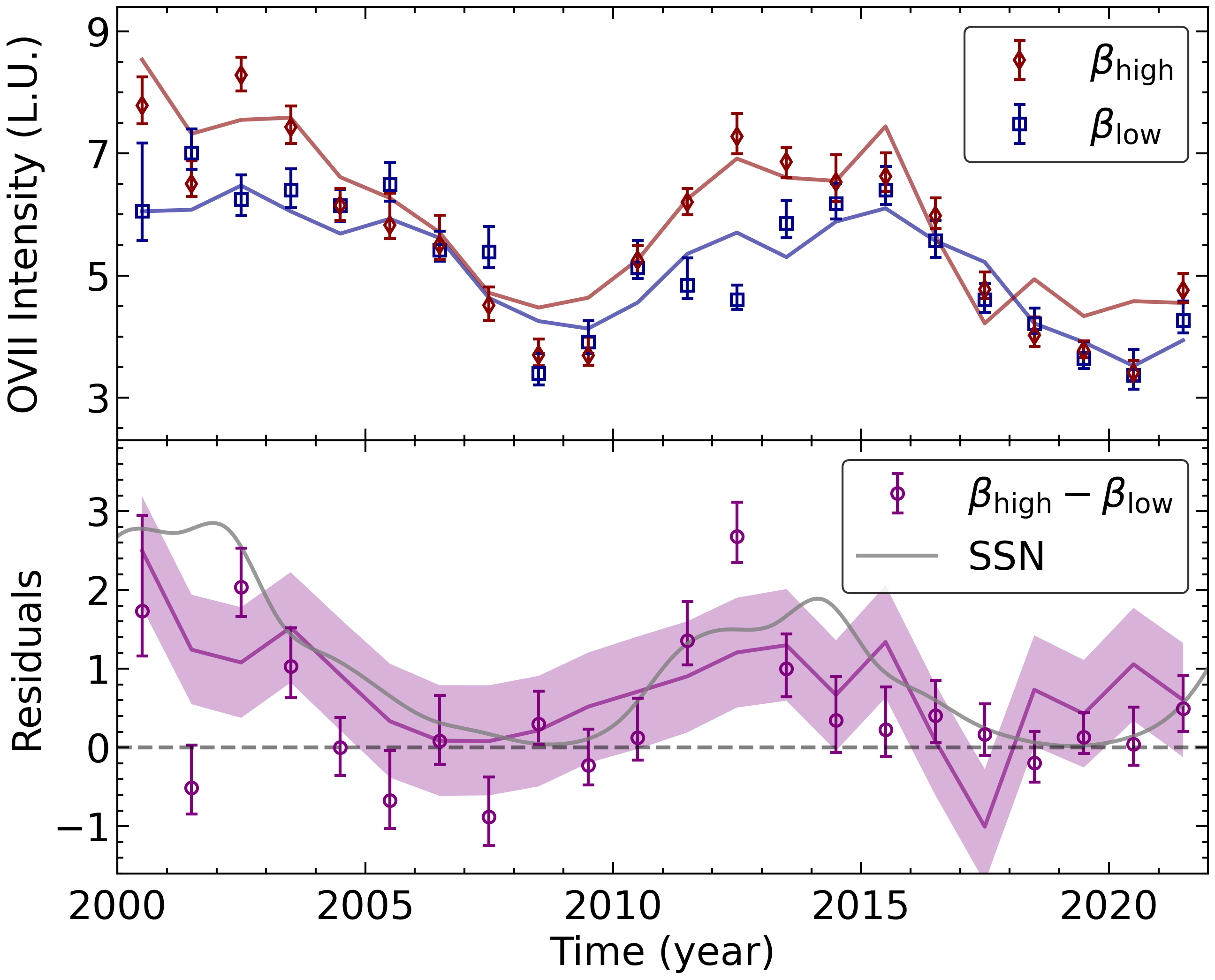}
\caption{\ion{O}{7} intensity variations in ecliptic latitude. Top: median \ion{O}{7} intensities derived from ``rebin'' and ``close-pair'' methods in low and high ecliptic latitudes ($|\beta|<30\degr$ and $|\beta|\geq30\degr$) over 1-year bins. Bottom: the intensity differences and $1\sigma$ uncertainty from the ``close-pair'' result, along with smoothed SSN. The \ion{O}{7} intensity is generally higher at higher ecliptic latitudes, especially during solar maxima. This variation likely results from the varying solar wind properties at different latitudes.}

\label{fig:ecliptic_variation}
\end{figure}

\subsection{Spatial Dependence of SWCX on Ecliptic Latitude}
Apart from the temporal variations, SWCX also exhibits spatial variations \citep[e.g.,][]{fujimoto2007evidence, ringuette2023observations}. 
Here, we investigate its correlation in ecliptic latitude by dividing the clean sample into low ($\beta_{\rm low} \equiv |\beta|<30\degr$) and high latitude ($\beta_{\rm high} \equiv |\beta|\geq 30\degr$) sub-samples.
In the top panel of Figure \ref{fig:ecliptic_variation}, \blue{we show the 1-year re-binned \ion{O}{7} intensities for these sub-samples, along with their best-fit models derived from the ``close-pair'' method outlined in Section \ref{sub:Helio SWCX Method}. }
The lower panel presents the differences between the two sub-samples and the solar cycle, traced by the SSN.
The difference suggests that \ion{O}{7} intensities are higher at high latitudes, especially during solar maxima. Conversely, during periods of low solar activity, the intensity difference diminishes, approximating zero.
Since the major differences occur during solar maxima, this latitude-dependent variation is likely attributed to the difference in solar wind properties across ecliptic latitudes.

Understanding the spatial distribution of SWCX is challenging due to multiple factors, including anisotropic solar wind and neutral gas distributions. Future research should aim to develop a more comprehensive model that incorporates these variables.

\subsection{Future Work}
\label{sec:FutureWork}
 
Considering the limitations in the SWCX and LHB modeling, our future work will first focus on further refining the foreground correction, accounting for short-term and spatial variations in SWCX, among other potential factors.
An accurate foreground model is the basis for further study of the MW hot gas, including the temperature structures (also see \citetalias{Qu:2023aa}) and the density distribution.
The density distribution is essential for estimating the total hot gas mass in the MW \citep[e.g.,][]{miller2015constraining}, which can provide further insights into fundamental questions, such as the ``missing baryon'' problem and feedback processes.

\section{Summary} 
\label{sec:summary}
In this study, we introduce the X-LEAP program and report a new set of \ion{O}{7}, \ion{O}{8}, and Fe-L line measures. They are extracted from SXRB spectra based on 5418 EPIC-MOS observations conducted before 2022. These lines represent spectral features at energies of $\approx$ 0.56 keV, $\approx$ 0.65 keV, and $\approx$ 0.80 keV.
\blue{The dataset is specifically optimized to study diffuse emissions from both the MW hot gas and the SWCX, minimizing the contamination from point and extended sources, as well as from irrelevant backgrounds.}

The key findings are summarized as follows:
\begin{enumerate}
\item 
All line measures are summarized in a machine-readable table, which is available online.
\item 
In the dataset, the observed \ion{O}{7}, \ion{O}{8}, and Fe-L line intensities have medians of $\approx$ 6 L.U., $\approx$ 1 L.U., and $\approx$ 1 L.U., respectively, with 90\% falling within the ranges of $\approx 2$--18 L.U., $\approx 0$--8 L.U., and $\approx 0$--9 L.U. (Figure \ref{fig:I_dist}). The strong positive correlation between these intensities suggests consistent intensity ratios, in line with the CIE model's prediction of a narrow temperature range in MW hot gas. In addition, observations inside eRBs contribute most high-intensity measurements, while high-intensity measurements outside the eRBs point to the existence of small-scale structures in the MW hot gas.
\item
\red{Long-term variations are observed in the \ion{O}{7}, \ion{O}{8}, and Fe-L intensities with medians of 1.8, 0.3, and 0.1 L.U., respectively (Figure \ref{fig:four_var}). This indicates SWCX contributes about 30\%, 20\%, and 10\% on average to these observed intensities.}

\item We present all-sky maps of SWCX- and absorption-corrected \ion{O}{7}, \ion{O}{8}, and Fe-L intensities (Figure \ref{fig:ALLSKYmaps}). These maps represent the soft X-ray emissions of the MW hot gas at different bands. The oxygen emissions extend beyond the eRBs and decrease from the Galactic center. Conversely, the Fe-L emission closely traces the eRBs and the Galactic disk, implying two different origins.
\item The Al and Si instrumental line centroids shift similarly away from their rest energies, likely due to temperature variations in the EPIC-MOS. These temperature changes alter line energies across the entire MOS spectrum, thereby distorting the $E_{\rm OVII}$ measurements. After addressing this effect, the corrected $E_{\rm OVII}$ shows an anti-correlation with solar activity (Figure \ref{fig:EO7}). This suggests that $E_{\rm OVII}$ can serve as an independent tracer of SWCX contributions, even when the spectral resolution is lower than the splitting of the \ion{O}{7} triplet.

\item The \ion{O}{7} intensities vary significantly over hours or days across observations of AA Tauri and Mars (Figure \ref{fig:samedir}).
These variations correlate closely with the solar proton flux, suggesting that they are associated with the magnetospheric SWCX.
\item The SWCX exhibits spatial dependence, with \ion{O}{7} intensity being higher at high ecliptic latitudes, especially during solar maxima (Figure \ref{fig:ecliptic_variation}). This variation is likely due to changing solar wind properties at different latitudes. Particularly, it lowers the observed intensity at the anti-center region, as its ecliptic latitude is low ($\approx5\degr$). 
\end{enumerate}

\section*{acknowledgments}
We thank the anonymous referee for their valuable comments that significantly improved our work. The authors also thank Henggeng Han, Zexi Niu, and Zikun Lin for their helpful comments on this work. 
JFL acknowledges support from the NSFC through grant Nos. 11988101 and 11933004, and support from the New Cornerstone Science Foundation through the New Cornerstone Investigator Program and the XPLORER PRIZE.
JNB would like to acknowledge the University of Michigan for its support.
This research is based on observations obtained with \xmm, an ESA science mission with instruments and contributions directly funded by ESA Member States and NASA.

\bibliographystyle{aasjournal}

\end{document}